\documentclass[amsmath,amssymb,aps,prb,lengthcheck,superscriptaddress,twocolumn]{revtex4}
\usepackage[T1]{fontenc}
\usepackage[latin1]{inputenc}
\usepackage{amsmath}
\usepackage{graphicx}
\usepackage{epsfig}
\usepackage{bm}
\usepackage{dsfont}
\usepackage{color}
\usepackage{braket}%simple ket{} and bra{} 
\usepackage{bbm}
\usepackage{physics}
\usepackage{comment}
\usepackage{mathtools}
\usepackage{multirow}
\usepackage{tabularx}

\begin{document}

\title{Loop current states and their stability in small fractal lattices of Bose-Einstein condensates }

\author{Georg Koch}
\affiliation{Institut f\"ur Physik, Universit\"at Greifswald, 17487 Greifswald, Germany}

\author{Anna Posazhennikova}
\affiliation{Institut f\"ur Physik, Universit\"at Greifswald, 17487 Greifswald, Germany}
\email[Email: ]{anna.posazhennikova@uni-greifswald.de}

%\date{\today}

\begin{abstract}

We consider a model of interacting Bose-Einstein condensates on small Sierpinski gaskets. We study eigenstates  which are characterised by  cyclic supercurrents per each triangular plaquette ("loop" states). For noninteracting systems we find at least  three classes of loop eigenmodes: standard; chaotic and periodic. Standard modes are those inherited from the basic three-site ring of condensates with phase differences locked to $2\pi/3$.  Standard modes become unstable in the interacting system but only when the interaction exceeds a certain critical value $u_c$. Chaotic modes are characterised by very different circular currents per plaquette, so that the usual symmetry of loop currents is broken. Circular supercurrents associated with chaotic modes  become chaotic for any finite interaction, signalling the loss of coherence between the condensates.  Periodic modes are described by alternating populations and two different phase differences. The modes are self-similar and are present in all generations of Sierpinski gasket. 
 When the interaction is included, the circular current of such a mode becomes periodic in time with the amplitude growing linearly with the interaction. Above a critical interaction the amplitude saturates signalling a transition to a macroscopic self-trapping state originally known from a usual Bose Josephson junction.  We perform a systematic analysis of this rich physics. 

\end{abstract}

\maketitle

\section{Introduction}

Fractal lattices are structures of non-integer dimension which are characterised by scale invariance rather than translational invariance of a typical Euclidean lattice \cite{MAN83}. In the context of solid state physics Sierpinski gasket lattices are particularly popular because their statistical mechanics properties and transport can be calculated analytically \cite{Gefen1981}. Their one-particle spectrum and density of states can be found by a relatively simple decimation procedure, resulting in discrete mostly degenerate eigenenergies with underlying fractal properties \cite{Domany1983,Rammal1984}. Corresponding eigenstates 
 are also nontrivial, with many of them being localized just due to the gasket geometry, i.e. in the absence of any disorder. 

Recently, a number of experimental successes in realisation of fractal structures \cite{Shang2015,Fan2014,Zhang2016,Kempkes2019,Liu2021,Biesenthal2022}
have been achieved. This has revived the interest in fractals, in particular,  within the theoretical condensed-matter community  with the latest publications covering a broad range of topics, including electron transport \cite{Katsnelson2016,Katsnelson2017,Katsnelson2018}, localisation \cite{Pal2012,Manna2021}, topology \cite{Neupert2018,Fritz2020,Pai2019,Manna2022}, flat bands \cite{Pal2018} and quantum phase transitions \cite{Krcmar2018,Xu2017}, to mention a few. 

In this work we consider a system of weakly-linked interacting Bose-Einstein condensates (BECs) placed on a small fractal lattice, which is in essence a mean-field description of the Bose-Hubbard model on the lattice.  The advantage of using BECs is their high tunability and unprecedented quantum state control (see for instance \cite{Dupont2021} and references therein). Since we study small systems, our goal is not only to understand the system at hand, but also to answer the old but still important question whether one can infer information on the physics of  larger systems from a study of their finite-sized counterparts. We show, that this is indeed possible to some degree for given systems. 

If two BECs are weakly linked, there is an oscillating current between them due to the Josephson effect \cite{Smerzi1997}. When three condensates are joined in a ring, then Josephson effect gives rise to a cyclic supercurrent along the ring, provided phase differences between neighboring sites are fixed to $2\pi/3$ and all site populations are the same \cite{Tsubota2000,Paraoanu2002}. An $N_s$-site ring would need the phase differences equal to $2\pi/N_s$ in order to maintain a cyclic supercurrent. Such cyclic currents are interesting per se since they represent  topological defects (vortices) and are relevant for the Kibble-Zurek scenario and quantum phase transitions \cite{Tsubota2000,Paraoanu2002,Zurek2002}.  Rings of condensates supporting stable circular flows were recently realised experimentally with polariton condensates \cite{Cookson2021}.

When small rings of condensates are arranged into a regular lattice, one would expect under certain conditions circular currents per plaquette, exemplifying loop current lattice states. Loop current states are quite popular in strongly-correlated electron systems where they are related to orbital magnetism and are signatures of exotic states of matter, e.g. exotic superconductivity  \cite{Varma1997,Haldane1988,Mielke2022}. In lattices of neutral bosonic systems synthetic magnetic fields were realised giving rise to circular plaquette currents corresponding to fluxes with tunable values  \cite{Ketterle2013,Bloch2013}.

\begin{figure}
    \centering
    \includegraphics[width=1.0\linewidth, keepaspectratio]{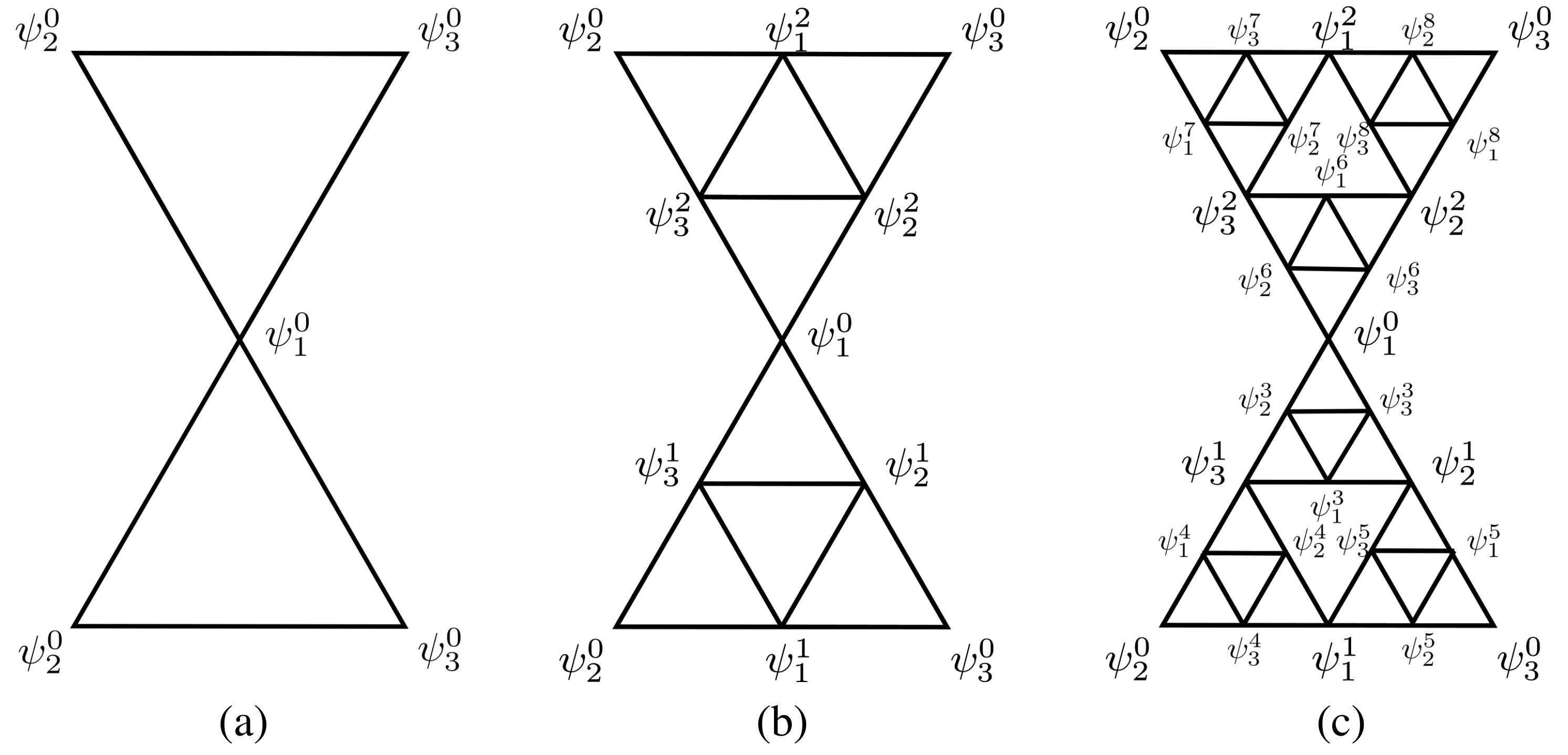}
    \caption{First three stages of the Sierpinski gasket fractal: (a) $k=0$, (b) $k=1$, and (c) $k=2$.  The wave functions on different nodes are denoted by an upper index representing the sub-triangle they belong to and a lower index which differentiates between sites within the subtriangle. (a) In  the first stage of the mirrored gasket, since the mirrored sites are set to be equal they have the same upper index. (b) In the middle stage two subtriangles have been added; and (c)  for the third stage  another six have been inserted. The sites denoted by the same wave-functions are the same sites. This is a result of the juxtaposition of the two largest triangles in order to secure periodic boundary conditions. }
     \label{fig:fracpic}
\end{figure}

In our work we consider a simpler case of a lattice with deep potential wells accommodating mini BECs with imprinted phases. The fact that the lattice has a fractal geometry makes the problem non trivial, since it is not a priori clear whether such a formation would support circular plaquette currents due to a  complicated discrete structure of the single-particle spectrum. Another question is, when it does, whether those currents remain stable in an interacting system. 
In the following we aim to answer these questions in  detail for the first three generations of the Sierpinski fractal. The Sierpinski fractal is generated from an equilateral triangle, recursively subdivided into smaller equilateral triangles.The first three stages of such subdivisions, taking into account periodic boundary conditions, are shown in Fig. \ref{fig:fracpic}.   

 With the help of a decimation procedure developed previously by the authors of Refs. \cite{Domany1983,Rammal1984} we show that loop states can indeed be realized in Sierpisnki gaskets of BECs, and in several different ways. For clarity, we divide them into  
three classes, all associated with various excited states. The ground state can only be described by equal site populations and zero phase differences between all sites and is therefore not looped. 

The three classes differ not only in realization of loop modes, but also in the way the looped states dynamics changes  in the presence of an interaction. For example, 
 {\it standard} modes are in many ways similar to the afore-mentioned modes of three-site condensate rings. Specifically, there is a critical interaction $u_c$ at which the corresponding fixed point in the phase space changes from stable to unstable. The circular current stays the same after the interaction is turned on, but for $u>u_c$ becomes chaotic upon a tiny perturbation. Interestingly, the critical interaction $u_c$ is different in each generation and  decreases with increasing fractal size. 

 The other class of loop eigenstates is termed by us {\it chaotic}. Chaotic eigenmodes are characterised by broadly distributed phase differences as well as population differences. All plaquettes have different circular currents, so that the usual symmetry of loop currents is broken. These modes do not survive in an interacting system, because any circular current associated with such a mode immediately acquires chaotic dynamics. 

 Another class of modes, which we refer to as {\it periodic}, comprises states deep inside the largest gap of the spectrum.  In contrast to chaotic modes, periodic states are represented  by only two phase differences and only two population differences at any stage. As a result, the phase portrait of the multidimensional noninteracting fixed point has effectively only two subspaces: stable and unstable. This fixed point changes when  the interaction is on, but the fundamental structure of the phase portrait remains the same. 
 
When  the interacting system is initially at the noninteracting fixed point, circular currents begin to oscillate in time. The character of these oscillations depends on the interaction. For sufficiently small interaction the amplitudes of the oscillations grow linearly with time. When the interaction exceeds a certain value $u_{ST}$, the amplitude saturates. The interaction $u_{ST}$ marks thus a transition of the system to a macroscopically self-trapped state similar to the one well-known from   
 the usual bosonic Josephson junctions \cite{Smerzi1997}. 
 
 We also find loop states at the gap edge, which have 
 another unusual property: They only have supercurrents along their inner edges, effectively dividing the system into two independent subsystems. Their basic structure remains the same with further generations as we explain in Appendix B. 

 The paper is organized as follows: in Sec. \ref{model} we briefly discuss the  model and equations of motion. In Sec. \ref{sec:nonint} we consider noninteracting fractals and their loop eigenstates. For convenience of the reader, we divide this section into three subsections:  Sec. III A briefly outlines known results of an infinite Sierpinski gasket, Sec. III B details loop eigenmodes of stages $k=0$ and $k=1$, Sec. III C deals with all the loop modes of stage $k=2$. Many analytical details are in Appendix B. In Sec. \ref{sec:int} we scrutinise the dynamics of the found eigenstates once the interaction between bosons in condensates is taken into account. We subdivide this Section into four subsections depending on the class of loop modes, since they exhibit very different dynamics. We summarise our work and discuss conclusions in Section \ref{concl}. 

%%%%%%%%%%%%%%%%%%%%%%%%%%%%%% MODEL %%%%%%%%%%%%%%%%%%%%%%%%%%%%%%%%%

\section{Model and equations of motion}

\label{model}

We consider a system of $N$ interacting bosons which are trapped in a fractal-shaped potential with the number of sites $N_k$ depending on the generation $k$ (specifically, we consider the first three generations of the Sierpinski gasket shown in Fig. \ref{fig:fracpic}). Generally, bosons on a lattice are described by the Bose-Hubbard Hamiltonian \cite{Fisher1989}. For a weak on-site interaction between bosons $U$ and macroscopic site occupancies $\rho=N/N_k$, a mean-field approximation can be justified \cite{Engl2014,Dubertrand2016,Mauro2021,Richter2022}. 
The system is then described semi-classically by a condensate wave-function $\Psi=(\Psi_1,\Psi_2,\dots, \Psi_{N_k})$, where the $\Psi_i$ obey  
the discrete nonlinear Schr\"odinger equation
\begin{equation}
 i \hbar\frac{\partial}{\partial t}\Psi_i(t)= U\vert \Psi_i(t)\vert^2\Psi_i(t)-K\sum_{<j>}\Psi_j(t) 
 \label{bas_schr}
 \end{equation}
 where $K$ describes tunneling between nearest neighbors.  This is equivalent to Gross-Pitaevskii equations with spatial dependencies integrated out, which works surprisingly well even for a two-well potential \cite{Ananikian2006}. 
 
 Note that in the description of equations we use a one-index site notation for simplicity, where possible. The sum in \eqref{bas_schr} goes over four nearest neighbours of site $i$. Condensate wave-functions are time-dependent complex functions 
 \begin{equation}
   \Psi_i(t)=\sqrt{N_i(t)}e^{i\theta_i(t) },
   \label{wave_f}
 \end{equation}
 where $N_i$ is the number of particles in the condensate on site $i$, and $\theta_i$ is the local phase. 
 Since the system under consideration is closed, the total particle number  $N=\sum_i N_i(t)$ is conserved. It is convenient to normalise the wave -functions \eqref{wave_f} by the root of the filling factor $\rho$ as 
 \begin{equation}
     \psi_i=\frac{\Psi_i}{\sqrt{\rho}}=\sqrt{n_i(t)}e^{i\theta_i(t) }.
 \end{equation}
 In this case  $\sum_i n_i=N_k$. 

Assuming $\hbar\equiv 1$  and introducing a dimensionless interaction constant 
 \begin{equation}
     u=\frac{U\rho}{K}
 \end{equation}
 we get from \eqref{bas_schr} the  discrete nonlinear Schr\"odinger (DNLS) equation
\begin{equation}
 i\frac{\partial}{\partial t}\psi_i(t)= u\vert \psi_i(t)\vert^2\psi_i(t)-\sum_{<j>}\psi_j(t).
 \label{fin_schr}
 \end{equation} 
We should mention here, that DNLS equations have a long history, with many examples of nontrivial as well as chaotic behaviour already in one-dimensional systems (see e.g., works in the 1980s-1990s \cite{Kenkre1986,Wan1990,Kalosakas1994,Hennig1995}), and also many recent achievements summarised in Ref. \cite{kevrekidis2009discrete}. Here we essentially investigate the complex domain of DNLS equations on fractal lattices.
 
 We make use of periodic boundary conditions, employed in \cite{Domany1983}, in order  to calculate the spectrum of the one-particle Schr\"odinger equation on such lattices. The boundary conditions utilise a mirrored gasket obtained by the juxtaposition of two identical generation $k$ gaskets at their corresponding external sites. In this way each site has exactly four neighboring sites.  

 Instead of complex equations \eqref{fin_schr} we can also solve a set of real equations for conjugated variables $n_i$ and $\theta_i$
 \begin{equation}
\begin{aligned}
    \dot{n}_i &= -2 \sum_{<j>}\sqrt{n_i n_j}\sin{(\theta_j-\theta_i)}, \\
    \dot{\theta}_i&= -u n_i + \sum_{<j>} \sqrt{\frac{n_j}{n_i}} \cos{(\theta_j-  \theta_i)}.
\end{aligned}
\end{equation}

We numerically evaluate the following quantities: pairwise phase differences between condensates on neighboring sites, pairwise population imbalances between condensates on neighboring sites, and circular currents. We first define the Josephson supercurrent between two 
adjacent condensates in the standard way 
\begin{equation}\label{eq:flux}
    I_{i,i+1}^{m,l}=2 \operatorname{Im}(\psi_i^{m*}\psi_{i+1}^{l}). 
\end{equation}
Here we had to revert to our double-indexed notation explained in Fig. \ref{fig:fracpic}, for clarity. The circular current per chosen subsystem is defined as a sum over  the Josephson subcurrents over all the structure
 divided by their number $N_I$ 
\begin{equation}
    I=\frac{1}{N_I}\sum_{i=1}^{N_I}\sum_{m,l}I_{i,i+1}^{m,l}.
\end{equation}
Generally we consider only the total circular current of the lower half fractal (summed over all the nodes). 
When the system under consideration  is in one of its eigenstates, then this circular current is constant, as well as all the phase differences and population imbalances.  We derive eigenstates analytically in Sec. \ref{sec:nonint} and verify numerically that circular currents remain indeed constant with time. 

This time-independent behaviour changes, however, if we add non-zero $u$ in our analysis, while using the eigenmodes as initial conditions. The resulting dynamics can be evaluated only numerically and reveals very different behavior depending on the initial conditions, i.e., on eigenstate classes. The detailed study of interacting systems is presented in Sec. \ref{sec:int}.

\section{Loop current states of the noninteracting system}

\label{sec:nonint}

\subsection{Loop eigenmodes of $k=0$ and $k=1$ stages}
\label{sec:flux01}

In this section we derive eigenmodes of the noninteracting system for stages $k=0$ and $k=1$, which are characterised by loop currents. Eigenvalues for any stage are well known and were derived by a decimation procedure in Refs. \cite{Domany1983,Rammal1984} (see also a brief reminder in Appendix A). Note that periodic boundary conditions (see Fig. \ref{fig:fracpic}) were crucial for development of the decimation procedure and are therefore employed in our work as well.  

\begin{figure}
    \centering
    \includegraphics[width=0.7\linewidth, keepaspectratio]{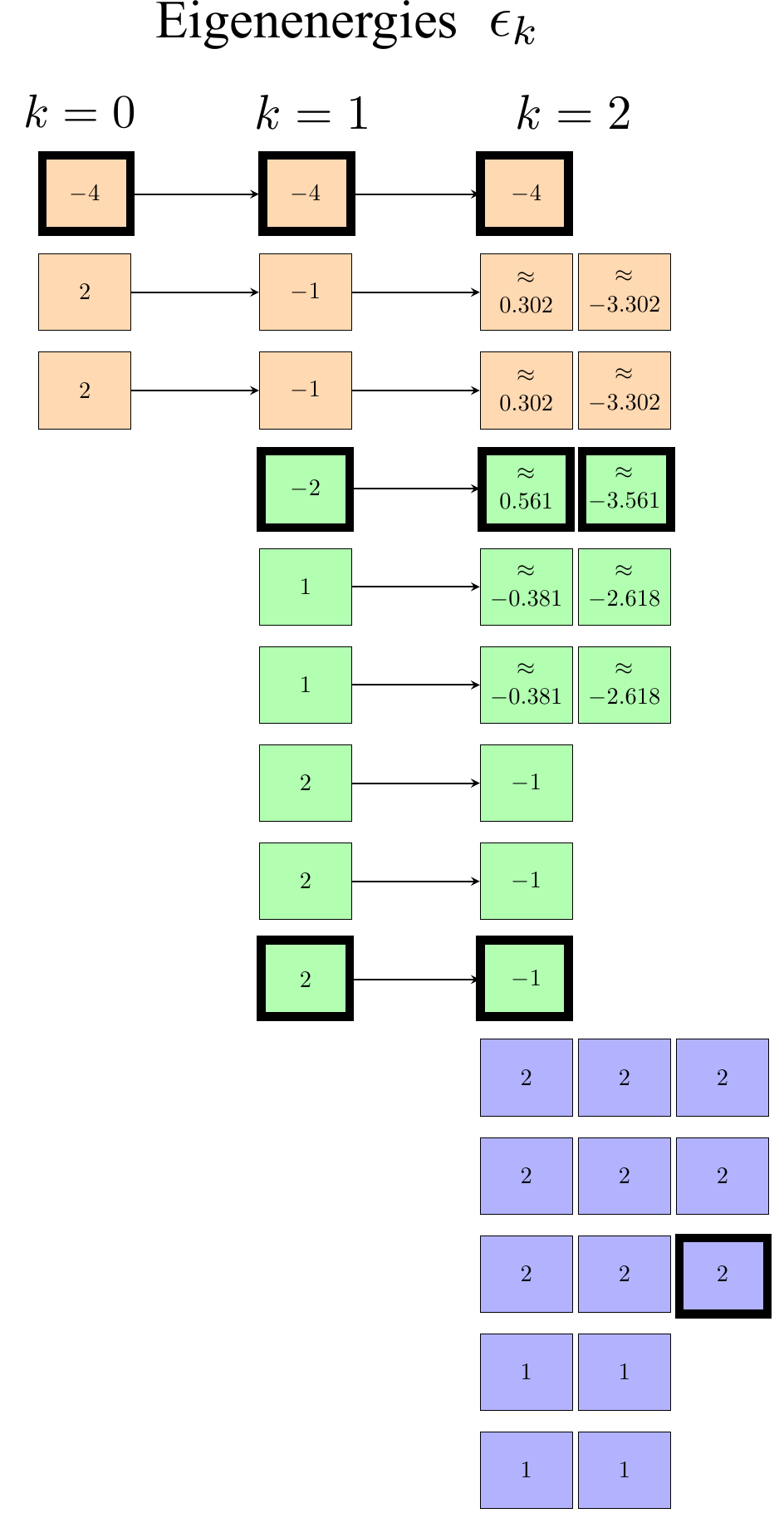}
    \caption{Flowchart of all eigenenergies (in units of $K$) for the first three generations of the Sierpinski gasket.  Values that have been generated    by the recursion procedure \eqref{eq:recursion} are connected by arrows with their ancestors. Those that are added on at a given stage without having direct predecessors are colored differently, e.g., the green values for $k=1$. 
     Thick black frames mark eigenvalues whose eigenvectors will not have any loop currents; hence such states will not be considered in this work. 
    }
    \label{fig:EE}
\end{figure}

The decimation procedure implies that some eigenvalues are derived from the others, and since this information is important for our derivations,  Fig. \ref{fig:EE} demonstrates a flowchart 
 where we present all eigenvalues of the first three generations and the relations between them. We also mark eigenvalues whose eigenstates we are not going to analyse since they can not maintain loop currents. 

 For stage $k=0$ the Hamiltonian (expressed in the units of  the coupling constant $K$) is trivial
\begin{equation}
    \mathcal{H}_{k=0}=
    \begin{pmatrix}
         0& -2& -2 \\
         -2& 0 & -2 \\
         -2& -2& 0 \\
    \end{pmatrix},
        \label{eq.frac_0}
\end{equation}
with eigenvalues $\epsilon_{k=0}\in\{-4,2,2\}$. The eigenmodes $|\Psi_{k=0}(\epsilon_{k=0}) \rangle$ are also readily found 
\begin{eqnarray}\label{eq.eigenk00}
    \ket{ \Psi_{k=0}(-4)}&=&
    \begin{pmatrix}
         1\\
         1\\
         1\\
    \end{pmatrix}, \\
   \ket{ \Psi_{k=0}(2)}_{\pm}&=&
   \begin{pmatrix}
        1\\
        e^{\pm i2 \pi/3}\\
        e^{\pm i4  \pi/3}
   \end{pmatrix}.
\end{eqnarray}
where the loop eigenmodes correspond to the doubly degenerate level $\epsilon_{k=0}=2$. These kind of states are well known from studies of rings of condensates  \cite{Tsubota2000,Paraoanu2002,Wozniak2022}. The state $\ket{ \Psi_{k=0}(-4)}$ corresponds to the ground state and is the same in all generations of the Sierpinski gasket (with the size increasing correspondingly). 

Since we use one of the loop eigenvectors for the derivation of loop states on stages $k=1$ and $k=2$, we introduce the shortened notation
\begin{equation}
    V\equiv  \ket{ \Psi_{k=0}(2)}_{+}.
    \label{eq:V}
\end{equation}

At the next stage $k=1$ the system  Hamiltonian still has a rather simple structure
\begin{equation}
     \mathcal{H}_{k=1}=
    \begin{pmatrix*}[r]
         \hat 0 & \hat S & \hat S \\
         \hat S & \hat S & \hat 0 \\
         \hat S & \hat 0 & \hat S \\
    \end{pmatrix*}
    , \quad \hat S=
    \begin{pmatrix*}[r]
         0& -1& -1 \\
         -1& 0 & -1 \\
         -1& -1& 0 \\
    \end{pmatrix*}
    \label{eq:Ham_gen1}
\end{equation}
with eigenvalues $\epsilon_{k=1}\in\{-4,2,2,2,-1,-1,-2,1,1 \}$. Since we are only interested in loop eigenstates, and they always come in conjugated pairs, we only explore states corresponding to the degenerate eigenvalues $\{2,\pm 1\}$. 

Now we notice that $V$ is an eigenstate of $\hat S$ with eigenvalue $\lambda_s=1$. We can use this fact for finding loop eigenmodes of the Hamiltonian $\mathcal{H}_{k=1}$. We notice that the simplest $3\times 3$ anticirculant matrix representing the structure of the Hamiltonian reads
\begin{equation}
    \begin{pmatrix*}[r]
         0 & 1 & 1 \\
         1 & 1 & 0 \\
         1 & 0 & 1 \\
    \end{pmatrix*},
\end{equation}
and eigenvectors of this matrix are readily found: $(1,1,1)^T, (-2,1,1)^T$ and $(0,-1,1)^T$. Hence the loop eigenmodes of $\mathcal{H}_{k=1}$ are
\begin{eqnarray} \label{eq:EVk=1}
      \ket{ \Psi_{k=1}(2)}&=&\begin{pmatrix} V \\ V \\ V \end{pmatrix}, \quad   \ket{ \Psi_{k=1}(-1)}=\begin{pmatrix} -2V \\ V \\ V \end{pmatrix}, \nonumber \\
       \ket{ \Psi_{k=1}(1)}&=&\begin{pmatrix} 0 \\ -V \\ V \end{pmatrix}.
\end{eqnarray}
The first eigenvector corresponds to $\epsilon_{k=1}=2\lambda_s=2$, the second to $\epsilon_{k=1}=-\lambda_s=-1$ and the third one to $\epsilon_{k=1}=\lambda_s=1$. The sequencing of condensates wave-functions inside eigenvectors  proceeds according to the following notation:  $(\psi_1^0\, \psi_2^0\, \psi_3^0\, \psi_1^1\, \psi_2^1\,  \psi_3^1\, \psi_1^2\, \psi_2^2\, \psi_3^2)^T$. The eigenmodes \eqref{eq:EVk=1} along with their complex conjugates constitute the six loop eigenvectors of the $k=1$ system. The remaining three eigenvectors can be obtained in the same way with $V$ replaced by $\ket{ \Psi_{k=0}(-4)}$, but they do not maintain loop supercurrents  and we do not consider them here. 

\begin{figure}
    \centering
    \includegraphics[width=0.9\linewidth, keepaspectratio]{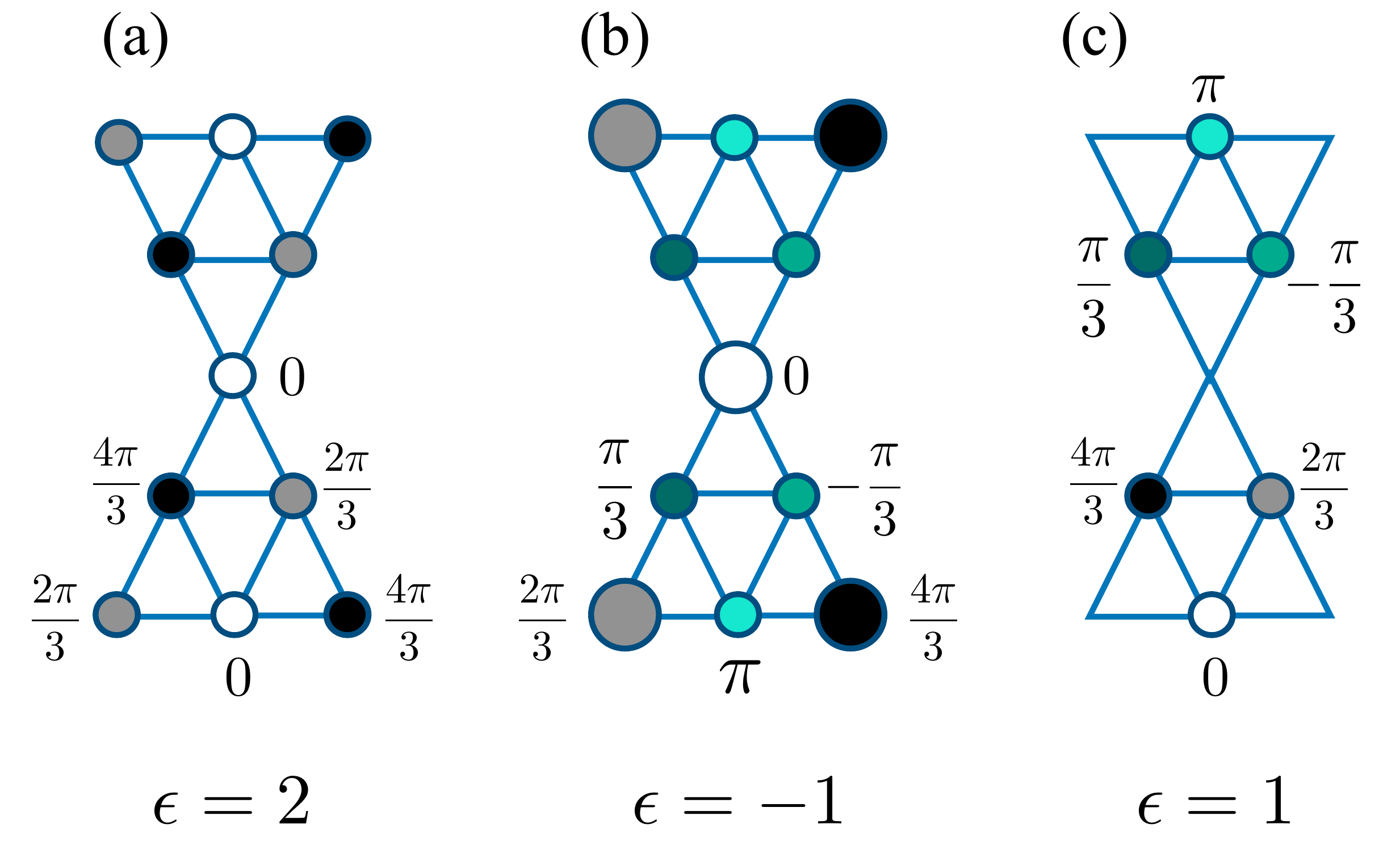}
    \caption{Schematic representation of the loop current eigenstates of stage $k=1$. A circle represents a Bose-Einstein condensate. Circles of the same size correspond to the same filling $\rho$, such that the filling of all sites in (a) is the same; however, in (b) some of the sites have larger filling factors, shown by larger circles; in (c) some sites are not filled at all. Different colors refer to different phases of condensates, for example, white means phase $0$; gray, phase $2\pi/3$; bright blue,  phase $\pi$; etc. 
    }
    \label{fig:EVk=1}
\end{figure}

Before proceeding to the $k=2$ case, we briefly discuss the results \eqref{eq:EVk=1}, schematically presented in Fig. \ref{fig:EVk=1}. The first eigenvector $\ket{ \Psi_{k=1}(2)}$ corresponds to the upper bound of the spectrum $\epsilon=2$, the energy value  with the largest multiplicity \cite{Domany1983} (see Fig. \ref{fig:DOS}). It is a trivial eigenvector since all triangles (large and small in Fig. \ref{fig:EVk=1}(a) are equivalent to the three-site ring with circulant current per triangle readily evaluated $I=\sqrt{3}$.  We also expect similar stability behavior as in three-site rings once the interaction is turned on (to be discussed in Sec. \ref{sec:int}). We refer to these modes as {\it standard} hereafter. 

The next loop state $ \ket{ \Psi_{k=1}(-1)}$ is schematically displayed in Fig. \ref{fig:EVk=1} (b) and  corresponds to the isolated energy state within the largest gap in the spectrum (see Fig. \ref{fig:DOS}). This mode 
is different from the standard mode in two ways: the inner triangles have smaller filling factors, expressed by smaller circles;  and phases of the BECs on the inner triangles are shifted by $\pi$ with respect to the standard configuration in Fig. \ref{fig:EVk=1}(a). We show later that this kind of modes will be present in all generations and we will refer to them as {\it periodic} for reasons explained in Sec. \ref{sec:int}.

Finally, we discuss states corresponding to the gap edge at $\epsilon=1$. In this case the Sierpinski gasket is only partially filled as is seen in Fig. \ref{fig:EVk=1} (c). The two inner triangles are then essentially uncoupled, but the phases in the upper inner triangle are shifted by $\pi$ with respect to the phases of the lower triangle. This is a partially looped state. 

\subsection{Loop eigenmodes of $k=2$ stage}

\label{sec:flux2}

The elegant way of deriving eigenvalues and eigenvectors presented in Sec. \ref{sec:flux01} can not be applied here due to the changed structure of the Hamiltonian; specifically, the Hamiltonian can not any longer be expressed solely in terms of $\hat S$-matrices. For this reason we resume with either the hierarchical derivation procedure \cite{Domany1983}, briefly outlined in  Appendix B, or with numerical calculations for special values. 

We start with the discussion of states corresponding to the spectrum's upper boundary $\epsilon=2$.  Out of the nine new cases (see the flowchart in Fig.  \ref{fig:EE}), eight can be potentially looped, which means only four states are of interest since loop eigenmodes come in conjugated pairs. One of the four vectors will be a standard eigenmode of the type shown in Fig. \ref{fig:EVk=1}(a). This is easy to demonstrate from the structure of the Hamiltonian (see Appendix B for details). The standard modes will  hence be  present in all generations of the Sierpinski gasket for $\epsilon=2$.

\begin{figure}
    \centering
    \includegraphics[width=1.0\linewidth, keepaspectratio]{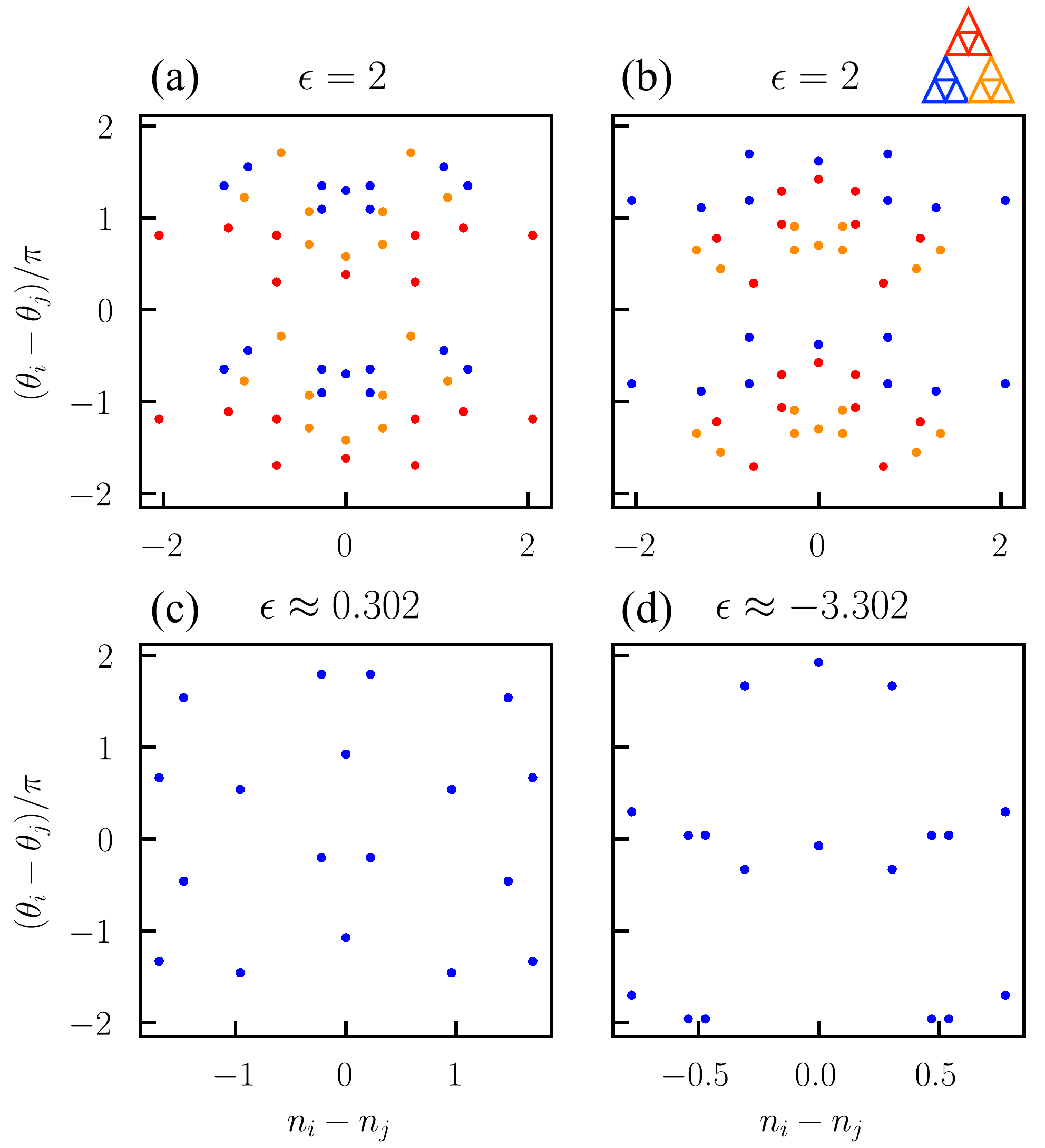}
    \caption{
    State space representation of {\it chaotic} eigenvectors in generation $k=2$ for different eigenenergies, indicated above the graphs. All phase differences of all nearest neighbors are plotted versus all population differences of all nearest neighbors. The small triangle on the top right explains the color coding within the fixed points corresponding to eigenvectors of (a) and (b). (a) corresponds to $|\psi_2(\epsilon=2)\rangle_2$,  (b) to $|\psi_2(\epsilon=2)\rangle_3$, (c) to $|\psi_2(\epsilon\approx -3.302)\rangle$ and (d) to $|\psi_2(\epsilon\approx 0.302)\rangle$ (see Appendix B). Blue, red and orange fixed points coincide for (c) and (d). 
    }
    \label{fig:EVk=2_eps2}
\end{figure}

The remaining three eigenmodes for $\epsilon=2$ are quite convoluted (see Appendix B), we refer to such modes as chaotic and choose to represent them in state space for clarity. In Fig. \ref{fig:EVk=2_eps2} (a) and (b) we show the associated state spaces for two out of three chaotic modes (this is essentially a representation of the multidimensional fixed point). We color-coded them according to triangles the fixed points (or rather components of the multidimensional fixed point) belong to. The color-coding was motivated by the fact, that at this stage the blue, red and orange subtriangles are not equivalent any more in terms of their fixed points, and as a consequence circular currents will be also different. This kind of behaviour was not observed in stage $k=1$. Note also, that the mode in Fig. \ref{fig:EVk=2_eps2} (b), if complex conjugated, has the same structure as the one in Fig. \ref{fig:EVk=2_eps2} (a).

The last eigenvector from this series $|\psi_2(\epsilon=2)\rangle_4$ is the same as the one in Fig. \ref{fig:EVk=2_eps2} (a), but with colors interchanged: blue points turn red, orange ones turn blue and red ones turn orange. We remark that it is possible to retrieve a symmetric eigenmode by a linear combination of the three chaotic modes. The resulting state space can be seen in Fig. \ref{fig:SymmMode} in Appendix B.

Associated state spaces for chaotic modes are quite "busy" with fixed points distributed all over the space and population differences $n_i-n_j$ taking values between $0$ and approximately $ 2$.  In contrast, standard eigenmodes have just one fixed point, because all population differences are the same and equal to zero, whereas all phase differences are equal to $2\pi/3$ (see Fig. \ref{fig:EVk=1}(a)). 

We now proceed to modes derived from gap states of the previous stage, i.e. states corresponding to $\epsilon_{k=1}=-1$. Although the asymmetry between the sub-triangles is not present for them, i.e. fixed points for blue, orange and red triangles coincide,  the modes are very similar to the chaotic modes for $\epsilon=2$, in that there is a rather broad distribution of phase differences as well as populations differences (see Fig. \ref{fig:EVk=2_eps2} (c) and (d)). We hence place the modes in the chaotic class.

\begin{figure}
    \centering    \includegraphics[width=0.6\linewidth, keepaspectratio]{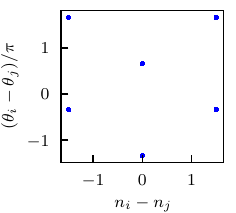}
    \caption{State space representation of {\it periodic} eigenvectors, stage independent ($k=1,2,...$): All phase differences of all nearest neighbors are plotted versus all population differences of all nearest neighbors.  
    }
    \label{fig:gapEV}
\end{figure}

 Periodic modes in generation $k=2$ are derived from the standard modes of generation $k=1$. It turns out (see Appendix B) that in terms of state space representation,  the new periodic modes are not different from the old ones: each new triangle will represent a standard ring, but with smaller occupation numbers compared to the triangle it is inserted in. The phases of the new triangle will be shifted by $\pi$ with respect to its hosting triangle as in Fig. \ref{fig:EVk=1}(b) and so on.  As a result the state space representation of such modes is  very simple and remains the same for any generation $k\ge 1$  (see Fig. \ref{fig:gapEV}). 

Finally, the partially filled modes of generation $k=1$ give rise to partially filled modes of generation $k=2$ with effectively separated subsystems with loop currents as we show in Appendix B. Since the dimensionless energies of these two new modes are irrational numbers (see the flowchart in Fig. \ref{fig:EE}), the modes acquire many broadly distributed fixed points within their corresponding subspaces.  Since this picture is not principally different from the chaotic modes, we do not elaborate on such modes in the following. 

All loop modes are characterised by constant circular current per plaquette, which can be calculated analytically or numerically. The values of the current can be very different; we give a number of examples for the total current  in Table \ref{table1}. 
\begin{table}[]
\centering
\caption{Total circular current $I$ (in units of $\rho K$) of the lower half fractal for stage $k=2$ and different energies $\epsilon$ (in units of $K$).}
\begin{tabularx}{3.4in}{||>{\centering\arraybackslash}X|>{\centering\arraybackslash}X||}
\hline
$\epsilon$ & $I$  \\ 
\hline
 $2$ (standard) & $\sqrt{3}$ \\
 $2$ (chaotic) & $0$ \\
  $\approx 0.302$ (chaotic) & $0.184$ \\
  $\approx -3.302$ (chaotic) & $0.104$ \\
 $-1$ (periodic) & $-\frac{\sqrt{3}}{2}$ \\
 \hline
\end{tabularx}
\label{table1}
\end{table}

We should also discuss edge states, i.e., states for $\epsilon=1$. They can not be derived by the decimation procedure and should be derived by other means. Naively, one would think that at each stage only the new subspace, i.e., only the smallest triangles, will be filled in such a way that a circular current will flow within. For example, at stage $k=2$ there will be six circular currents since there are six new triangles (see Fig. \ref{fig:fracpic}, $k=2$).  It turns out that this is only partially true (see Appendix B, after Eq. \eqref{eq:quasi_loc}). In fact, only four out of six triangles will be characterised by circular currents, whereas the remaining two will have only two sites out of three filled, e.g.  $(\psi_1^3,\psi_2^3,\psi_3^3)=(e^{-i2\pi/3},e^{i2\pi/3},1)$ (loop current), $(\psi_1^4,\psi_2^4,\psi_3^4)=(e^{i\pi},0,e^{-i\pi})$ (no loop current) and $(\psi_1^5,\psi_2^5,\psi_3^5)=-(1,e^{i2\pi/3},e^{-i2\pi/3})$ (loop current).

We thus derived all loop current states for stage $k=2$. For convenience, we summarise our results schematically in the next section. 

\subsection{Schematic summary}

\begin{figure}
    \centering
    \includegraphics[width=0.9\linewidth, keepaspectratio]{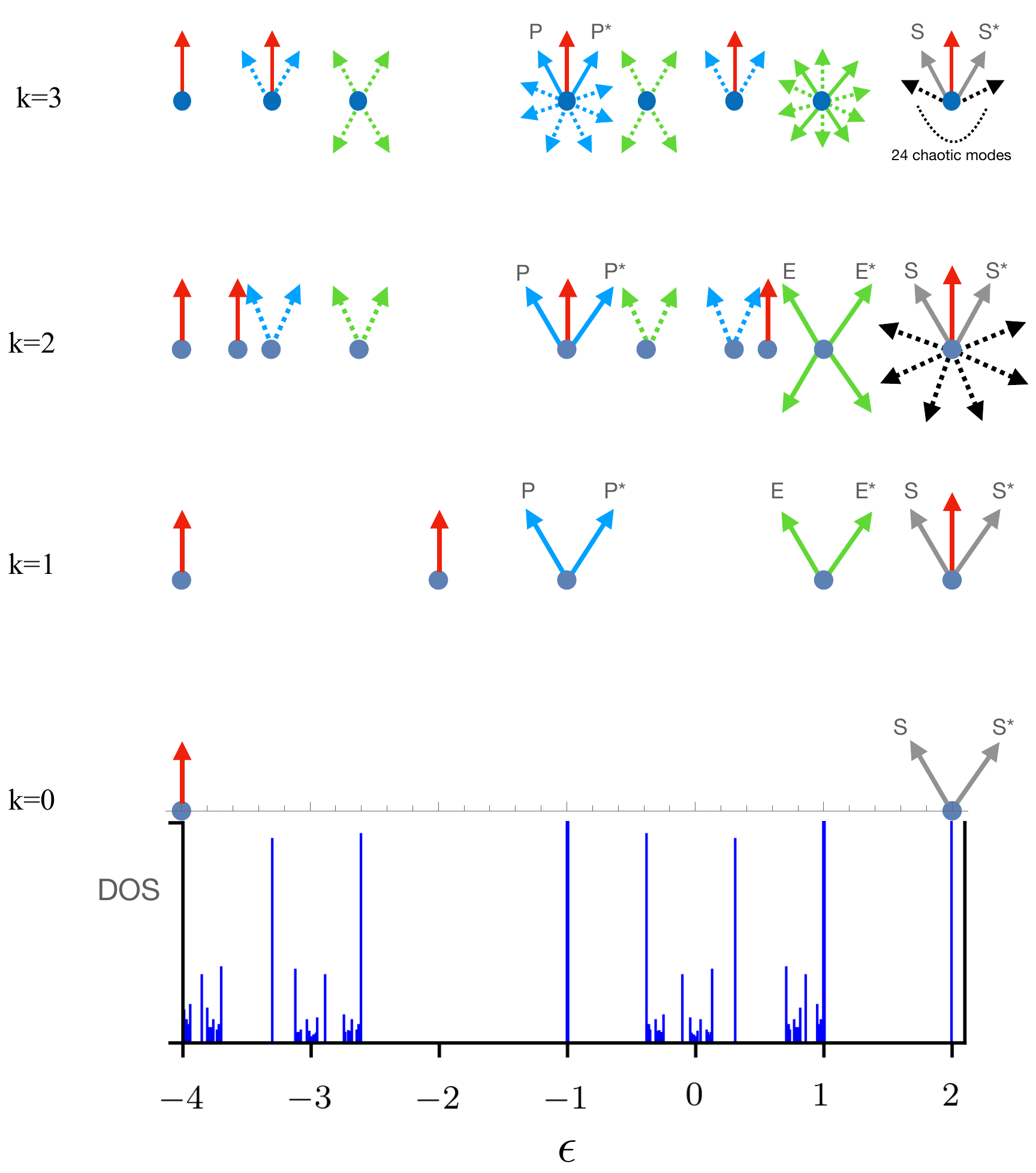}
    \caption{Schematic representation of all eigenstates of the first three generations of Sierpinski gasket ($k=0,1,2$) depending on energy (in units of $K$). A few  examples (not all eigenstates) are shown for $k=3$. Below the states DOS is appended, so that one can see whether a state is within the gap, at the gap edge or elsewhere.  Red (vertical) vectors represent real eigenstates. Tilted vectors represent loop current eigenstates.  "S" stands for standard mode, "P" for periodic and "E" for edge modes. Letters with stars mean complex conjugation. Dashed vectors indicate chaotic modes. }
    \label{fig:scheme}
\end{figure}

In Fig. \ref{fig:scheme} we show a resulting schematic diagram of all modes in generations $k=0,1,2$ and some of the modes in generation $k=3$. Modes that can not carry loop currents are marked by red vectors. Standard modes are colored gray and are present for all generations for $\epsilon=2$. With a degeneracy of $\epsilon=2$ increasing from generation to generation, the standard modes will be accompanied by chaotic modes, depicted by dashed vectors. 

Edge modes (these are the modes for $\epsilon=1$) are colored in green and are special, since they can not be derived by the decimation procedure, however, we managed to derive them by other means for $k=1$ and $k=2$. We assume their basic structure will be preserved for $k>2$. Edge modes in generation $k=1$ are ancestors of the pairs of chaotic modes (shown in green) in generation $k=2$. Note, that these pairs also correspond to band edges. The quadruple-degenerate edge mode in generation $k=2$ will give rise to chaotic (green) modes in generation 3 and so on.

Periodic modes (these are gap states for $\epsilon=-1$) are derived from standard modes of the previous generation and are present for all $k\ge 1$. For this reason they structurally resemble their ancestors. Periodic modes of generation $k=1$ give rise to pairs of gap modes in generation $k=2$ shown as blue dashed vectors. 

This procedure will continue ad infinitum. Gap states will give rise to gap states,  and partially filled edge states will give rise to partially filled edge states in the new generation. Inside each gap there will always be a real state, for example, the red state for $\epsilon=-2$. Such states will appear only once for each gap and are therefore not visible in the density of states plot. Nice, i.e., not chaotic loop current states will  only be present in each generation for $\epsilon\in\{-1,1,2\}$.  

\section{Inclusion of interaction and chaos onset}

\label{sec:int}

In this section we investigate the role of interaction $u$ in the behaviour of circular currents and stability of non-interacting fixed points representing loop eigenmodes discussed in the preceeding Section. The analysis is mostly done numerically by solving Eqs. \eqref{fin_schr} with the Runge-Kutta method. 

\subsection{Standard modes}

\begin{figure}
    \centering
     \includegraphics[width=0.99\linewidth, keepaspectratio]{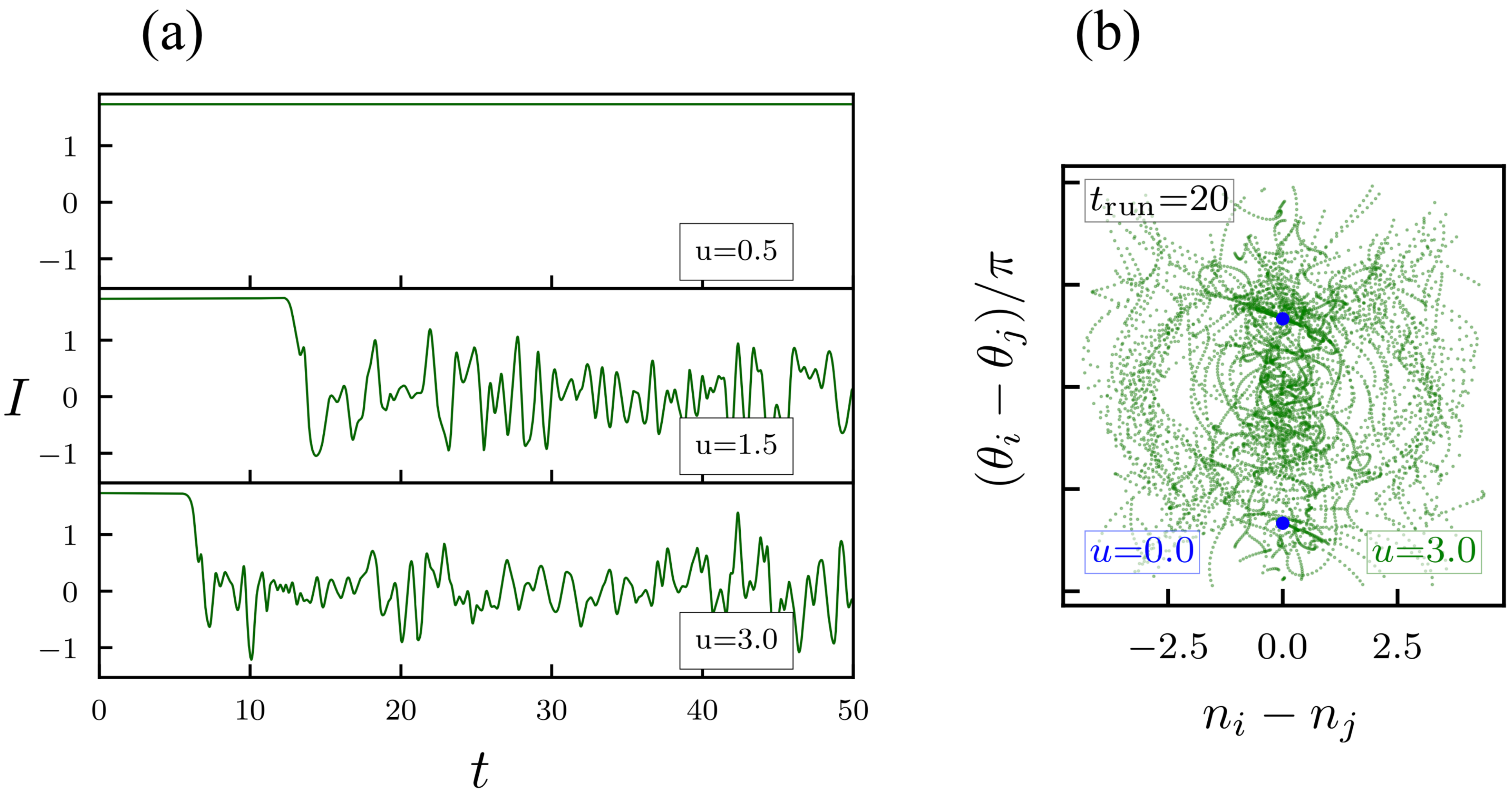}
           \caption{(a) Circular current $I$ (in units of $\rho K$) of the lower half fractal versus time (in units of $1/K$) for a standard mode of generation $k=1$ for different interaction values $u$. Random deviations of phases and occupations from their eigenmode values of the order of  $10^{-8}$ are introduced. (b) Phase space trajectories for the same mode after runtime $t_{\text{run}}=20$. For $u=0$ the mode is represented by fixed points shown by blue dots. For $u\neq 0$ the system becomes chaotic with trajectories covering all the phase space available. }
    \label{fig:Current_stand}
\end{figure}

Standard modes are different from all the other modes in that they remain fixed points of the DNLS equation  \eqref{fin_schr} even when interaction is turned on. This means that a circular current calculated along any triangular plaquette remains constant in time for $u\neq 0$. This interesting property is inherited from a standard three-site ring mode with equal populations and phase difference equal to $2\pi/3$ between all pairs of neighboring sites \cite{Tsubota2000,Wozniak2022}. The stability of the standard mode, however, depends on the interaction. In a ring the mode changes from stable to unstable at a critical value $u=u_c=1.5$, which can be derived analytically from linear stability analysis.  In the unstable regime the system can be forced out of the stationary state by the smallest perturbation and circular current plunges into chaotic dynamics, albeit not immediately, but after a period of time $t_c$ which depends on perturbation and interaction \cite{Wozniak2022}. We see this kind of behaviour in Sierpinski gaskets as well, exemplified in Fig. \ref{fig:Current_stand}. Fig. \ref{fig:Current_stand}(a) displays the circular current $I$ becoming chaotic with time depending on the interaction, whereas Fig. \ref{fig:Current_stand}(b) displays phase space chaotic trajectories associated with this behaviour. 

\begin{figure}[h]
    \centering
    \includegraphics[width=0.9\linewidth, keepaspectratio]{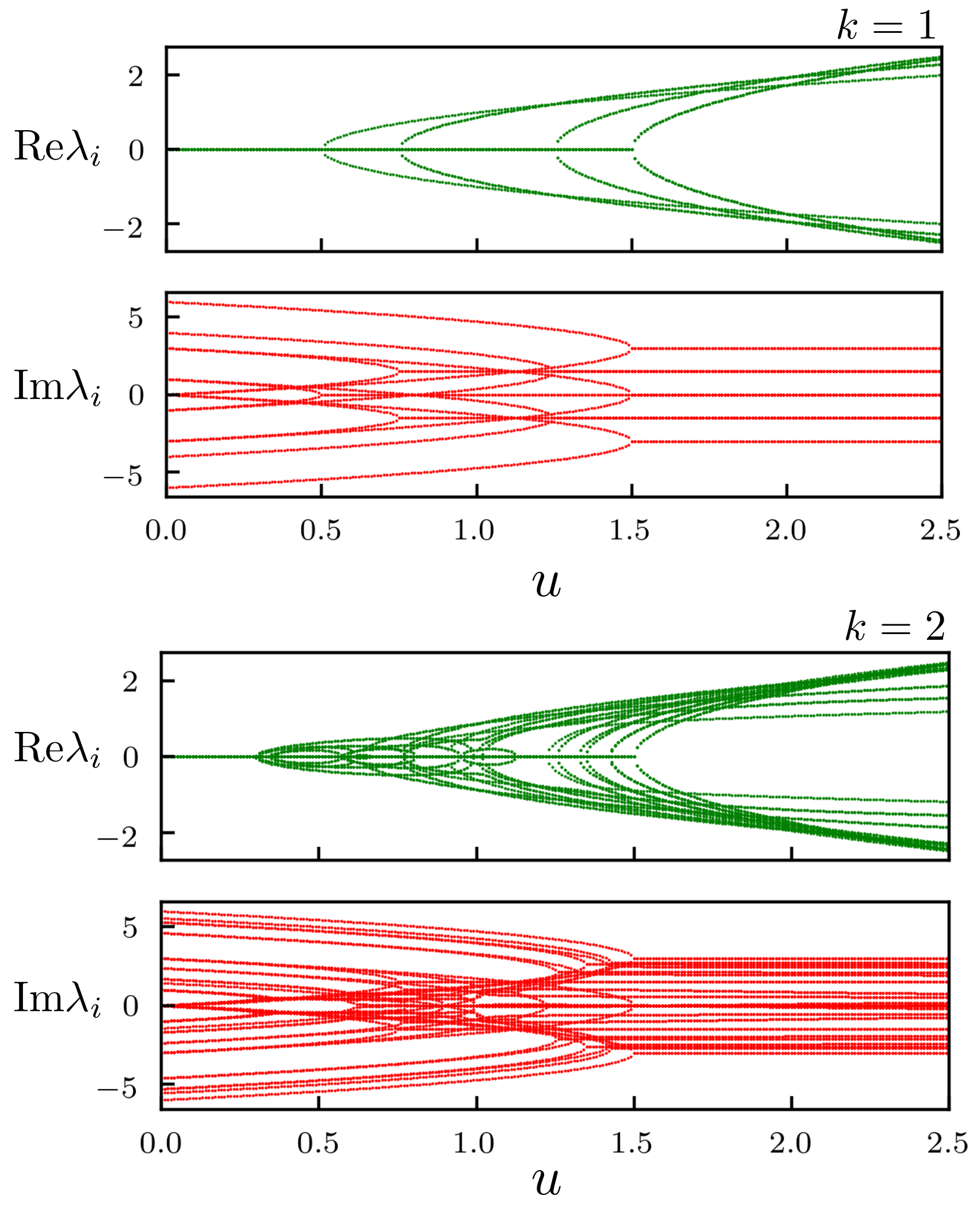}
     \caption{Real ($\Re \lambda_i$) and imaginary ($\Im \lambda_i$) parts of all Jacobian eigenvalues in units of $K$ versus  the dimensionless interaction parameter $u$, for the standard mode in generations $k=1$ and $k=2$. $u_c$ is determined by the first $\lambda_i$ acquiring a non-zero real part. }
    \label{fig:LinStab}
\end{figure}

To understand better this dynamics and to find $u_c$, we perform linear stability analysis of the standard modes for $k=1$ and $k=2$.  It is useful to convert to canonical coordinates first. All $n_i$ remain the same except for the one being eliminated though particle conservation, for example the $N$th coordinate $n_N=N-\sum_{i=1}^{N-1}n_i$. The corresponding phase variables are $P_i=\theta_N-\theta_i$ with $i=1,...,N-1$. The canonical variables satisfy the Hamilton equations of motion $\dot{n_i}=-\frac{\partial H }{\partial P_i}$ and $\dot{P_i}=\frac{\partial H }{\partial n_i}$. The initial condition remains a fixed point in these variables. The Jacobian $J$ of the size $2(N-1)$ then reads
\begin{equation}
    J=
\begin{pmatrix*}
    \frac{\partial \dot{n}_1}{\partial n_1} & \cdots & \frac{\partial \dot{P}_{N-1}}{\partial n_1}\\
    \vdots & \ddots & \vdots \\ 
    \frac{\partial \dot{n}_1}{\partial P_{N-1}} &  \cdots &\frac{\partial \dot{P}_{N-1}}{\partial P_{N-1}}
\end{pmatrix*}
\end{equation}
Calculation of its eigenvalues $\lambda_i$ is carried out numerically and the results for generations $k=1$ and $k=2$ are shown in Fig.  \ref{fig:LinStab}. We see that indeed, for small $u$-s all eigenvalues are purely imaginary and the system is stable, i.e. remains in its fixed point. However, once at least one eigenvalue acquires a non-zero real part, the stability is violated. We numerically identify the critical values of $u$ when this happens: $u_c(k=1)\simeq 0.51$ and $u_c(k=2)\simeq 0.30$. We see that $u_c$ indeed, depends on the generation. This is not very surprising, since the Jacobian is getting larger with each generation, which amounts to a broader distribution of its eigenvalues and therefore more complicated equations  for $u_c$. We expect, that in the thermodynamic limit $u_c$ will be tending to zero.  

\begin{figure}
    \centering        \includegraphics[width=0.9\linewidth, keepaspectratio]{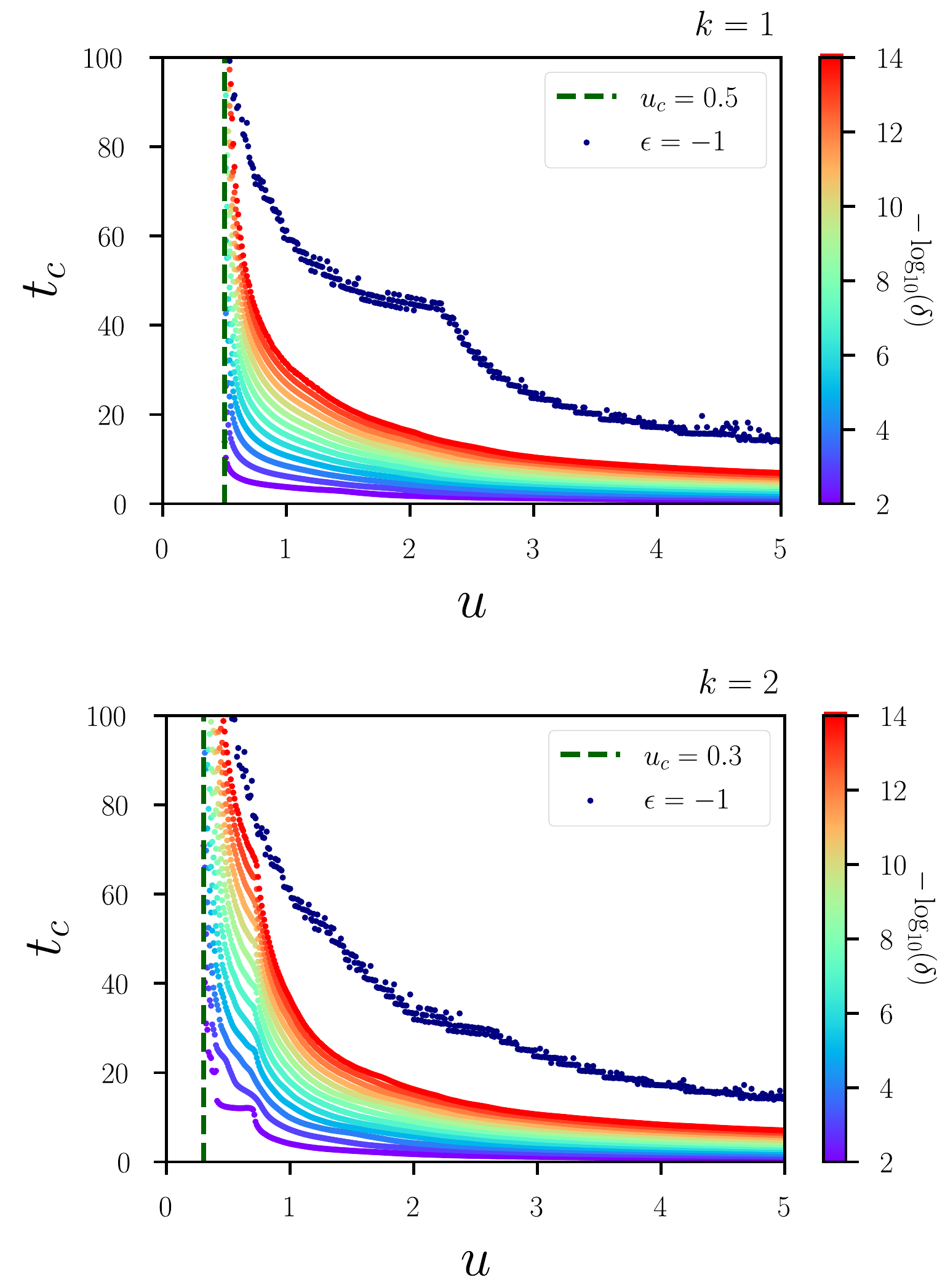}
        \caption{Critical time $t_c$ (in units $1/K$) dependent on  interaction $u$ for the standard mode of generations $k=1$ and $k=2$.  The scaling parameter $\delta$ of the initial phase and occupation displacement is varied from $10^{-2}$ to $10^{-14}$. The critical interaction derived from linear stability analysis is shown by the vertical dashed line for convenience. For comparison, $t_c$ of a periodic mode, i.e., a mode for $\epsilon=-1$, is shown with blue dots. }
        \label{Fig:TC}
\end{figure}

As discussed, the sliding into chaotic dynamics does not necessary happen immediately, so that there is a certain time-scale $t_c$ associated with that. In order to evaluate the time-scale we introduce small random deviations of the phases and occupation numbers of condensates according to 
\begin{equation}
\begin{aligned}
    &\theta_i \rightarrow \theta_i+\delta\, r_i,\\
    &n_i \rightarrow n_i + \delta\, r_j,
\end{aligned}
\end{equation}
where $r_{i,j}$ are random numbers within $[-1,1]$, and $\delta$ is a parameter.  We then numerically analyse $t_c$ dependent on $u$ for different values of $\delta$. The results are presented in Fig. \ref{Fig:TC}. As expected, $t_c$ diverges in the vicinity of $u_c$, found from linear stability analysis,   and decreases with the growth of $u$. If we fix $u$, then the characteristic time rapidly decreases  as the deviation $\delta$ from the stationary value becomes larger. 

\begin{figure}
\centering
\includegraphics[width=0.9\linewidth, keepaspectratio]{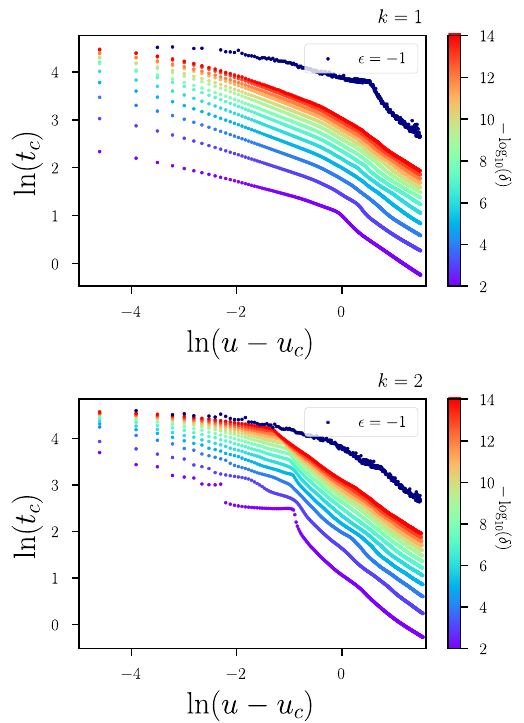}
    \caption{Double logarithmic representation of Fig. 8. The interaction parameter was shifted by $u_c$. Linear segments (before and after the kink) were fitted, with the resulting slopes presented in Tables \ref{slopes1} and \ref{slopes2}
     \label{fig:TC_LOG} of Appendix C.} 
\end{figure}

Interestingly, $t_c$ for the standard mode of generation $k=2$ has a kink at about $u\approx 0.7$. 
In order to better understand the $t_c$ behaviour, we show a doubly logarithmic plot of Fig.\ref{Fig:TC} in Fig.\ref{fig:TC_LOG}. We see that for both fractal stages there are two separate linear segments with different slopes (before the kink and after it). The kink is now identifiable even for $k=1$ around $u-u_c\approx 1$. The two stage behavior can also be seen in the $\epsilon = -1$ case, although less so for low values of $u$ at $k=2$. In general, this shows the power-law behavior of $t_c$ with respect to the interaction strength according to 
\begin{equation}
    t_c= (u-u_c)^ae^b.
    \label{eq:linear_fit}
\end{equation}
Here $a$ is the slope and $b$ the intercept in the respective segments. Tabulated values for the slope $a$  deduced from linear fitting of logarithmic plots can be found in Appendix C.

\begin{comment}
We attribute the kink to the unusual behaviour of the Jacobian eigenvalues $\lambda_i$ around this value of $u$. As can be seen in Fig. \ref{fig:LinStab} some of the real parts go back to zero around this value of $u$, thus trying to force the system back into a stationary state. This is reflected in the increased $t_c$ in the interval $u\in [u_c, 0.7)$ for $k=2$. 
\end{comment}

\subsection{Chaotic modes}

\begin{figure}[h]
    \centering
    \includegraphics[keepaspectratio]{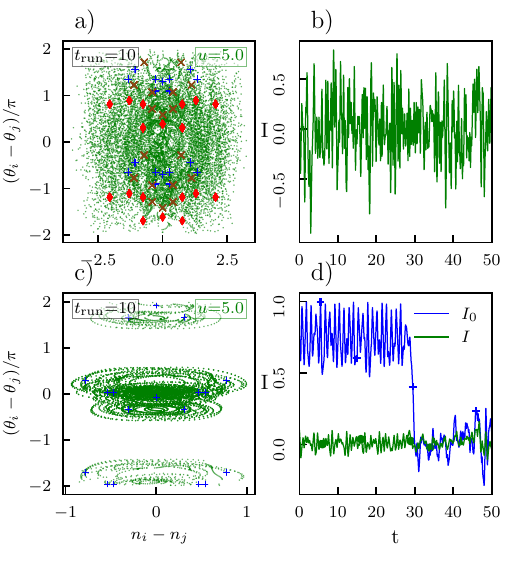}
     \caption{Chaotic modes for $u\neq 0$:
     (a) and (c) phase-space trajectories and (b) and (d) circular currents in units of $\rho K$ for energy (a) and (b) $\epsilon=2$ (in units of $K$), and  (c) and (d)  $\epsilon \approx -3.302$. The timerun  (in units of $1/K$) and interaction parameter are $t_{run}=10$ and $u=5.0$ for both cases. In (b) only total current $I$ is shown. In (d), apart from the total current (lower curve), also    the current along outer triangle $I_0$ only (upper curve) is presented. }
    \label{fig:chaotic_u}
\end{figure}

Chaotic modes shown in Fig. \ref{fig:EVk=2_eps2} appear only in the second generation of the Sierpinski fractal and are  characterised by components of their fixed point distributed all over the phase space. Since these modes are not fixed points of the interacting system, circular currents associated with them become immediately chaotic once the interaction is on. This behaviour is demonstrated 
 in Fig. \ref{fig:chaotic_u}, where we show one of the chaotic modes for $\epsilon=2$ (Figs. 11 (a) and 11(b)) and a mode for $\epsilon=-3.302$ (Figs. 11 (c) and 11(d)).

 Figs.11 (a) and 11 (c) depict phase space trajectories for $u=5$ after a certain runtime $t_{run}=10$. We see that in Fig. 11 (a) the system evolves all over the phase space, a behaviour characteristic of chaotic dynamics. In Fig. \ref{fig:chaotic_u} (b) the total current turns indeed chaotic for the interacting system signalling the immediate loss of coherence between BECs. 

 Figs. \ref{fig:chaotic_u} (c) and 11 (d) demonstrate that, interestingly,  the $\epsilon=-3.302$-mode first undergoes a weakly chaotic regime characterised by the circular currents oscillating around their stationary values. In the projected phase portrait this is reflected in semi-elliptic trajectories of the system close to some of the components of the fixed point (see Fig. \ref{fig:chaotic_u}(c)). 
Remarkably, this weak chaotic regime does not last forever but is associated with a time scale, similar to the standard mode behaviour in that it decreases with the growth of $u$.  We attribute this fascinating behaviour to the nonhomogeneous  distribution of the fixed point components in such a mode. Specifically,  we mean the quasi-isolated bunch of components concentrated in the narrow interval around $(\theta_i-\theta_j)/\pi=0$ and rather broadly distributed in $n_i-n_j$.  The time-scale $t_c$ is an effective  time period necessary for this quasi-isolated subspace to get connected with other subspaces, which inevitably happens when a system is chaotic. For example, for the case in Fig. \ref{fig:chaotic_u} d) $t_c\approx 30$. This time period gets shorter with increased interaction, as one would expect. 

\subsection{Periodic modes}

Finally, we discuss the periodic modes, i.e., modes corresponding to $\epsilon=-1$, which have the same state space representation in all generations for $u=0$. Similar to chaotic modes, these eigenstates do not remain fixed points in interacting system. However, cyclic supercurrents associated with these modes do not turn chaotic immediately either when $u\neq 0$. Instead, they undergo a temporary regime, where all supercurrents manifest periodic behaviour.  

\begin{figure}[h]
    \centering
    \includegraphics[ keepaspectratio]{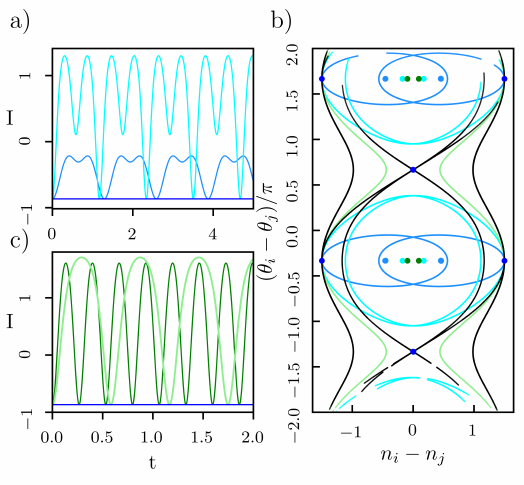}
     \caption{Periodic modes for $u\neq 0$. (b) Phase space trajectories for different values of $u$: $u=3$ (dark blue), $u=10$ (light blue), $u=12$ (black), $u=13$ (light green), and $u=20$ (dark green). Dots of the same color represent corresponding fixed points calculated from Eq. \eqref{fp_u}. The bright blue dots are the same as in Fig. \ref{fig:gapEV}. (a) and (c) show the currents associated with these interaction values  for $u<12.0$ (a), and for $u>12.0$ (b). There is a qualitative change at the self-trapping critical interaction $u_{ST}=12.0$. All currents (in units of $\rho K$) exclusively oscillate above the non-interacting value. The runtime for (a) is set to $t_{run}=5$ (in units of $1/K$) in order to not include the chaotic regime.  }
    \label{fig:periodic_dyn}
\end{figure}

We first examine this periodic dynamics for different values of $u$. In Fig. \ref{fig:periodic_dyn} (a) and (c) we show how the total current oscillations change with increasing $u$.  Interestingly, in all cases the current oscillates exactly above its noninteracting value, indicated by the straight blue line. The oscillations contain more than one frequency, which prompted us to perform Fourier analysis of the dynamics, which will be considered later in the text. 

In Fig. \ref{fig:periodic_dyn}(b) we demonstrate  the  phase portrait of our system prior to chaos onset. This is the plot for $k=1$; however, we obtained an identical graph for $k=2$ and hence assume this behaviour is typical of such modes in any generation. 
In Fig. \ref{fig:periodic_dyn} (b) one can notice that the dynamics crucially changes when $u$ exceeds a certain value $u_{ST}=12$. 
One can observe that for $u<u_{ST}$ the system progresses along closed almost elliptic trajectories. Closed trajectories mean that both phase differences and population differences oscillate around some values. This ceases to be the case when $u$ exceeds $u_{ST}$. Although population differences continue to oscillate around some nonzero values, the phase differences stop oscillating and turn instead into running phases. The transition to the running phase regime is termed self-trapping,  and is known from general systems of discrete nonlinear Schr\"odinger equations, as in,for example, Refs. \cite{Kenkre1986,Kalosakas1994}. In a system of two coupled Bose-Einstein condensates, the macroscopic self-trapped state was introduced in Ref.\cite{Smerzi1997} and experimentally verified in Ref. \cite{Albiez2005}.

\begin{figure}
    \centering
    \includegraphics{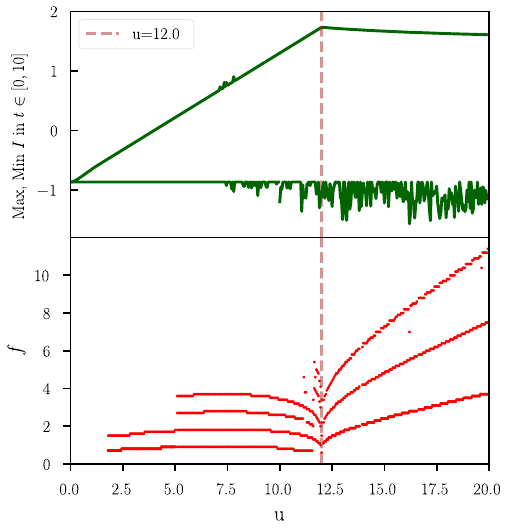}
    \caption{Upper panel: Maximum and minimum values of the total current (in units of $\rho K$) during $t_{\mathrm{run}}=10$ (in units of $1/K$) versus interaction  (for a periodic mode).
    Signs of chaotic behaviour are present for  $u>7.5$. Lower panel: Biggest frequency contributions to the circular current of periodic modes dependent on the interaction.  For the latter the frequencies of the largest detectable contributions to the Fourier transform are shown. Only values above $u=1.75$ are considered since the oscillations of the current below this value are very small and difficult to detect. Above $u=5.0$ two more frequency contributions can be detected. Around $u_{ST}$ there is noise due to difficulty in properly detecting frequencies.   } 
    \label{fig:AmpltPlots}
\end{figure}

 The transition to self-trapping at $u=u_{ST}=12$ is also evident from Fourier analysis of the total circular current presented in Fig. \ref{fig:AmpltPlots}. Prior to $u_{ST}$ the current is characterised by a multi-frequency dynamics with the frequencies only weakly dependent on $u$. Above $u_{ST}$ one can clearly distinguish three frequencies, which rapidly  increase with $u$. In the upper graph of Fig. \ref{fig:AmpltPlots} minimum and maximum  values of the total current dependent on $u$ in the time window $t\in [0,10]$. Although the minimum value stays interaction independent unless the system enters the chaotic regime, the maximum value is linearly-dependent on $u$ in the non-self-trapped case and almost constant or slightly decreasing in the self-trapped regime. Qualitative differences in oscillations and their Fourier transforms before and after self-trapping  were also found in bosonic Josephson junctions \cite{Raghavan1999}. 

 %%%%%%% NEW FIXED POINTS %%%%%%%%%%%%%

In addition to this analysis, we also find fixed points of the system when interaction $u$ is finite. As can be seen from the phase portrait in Fig. \ref{fig:periodic_dyn}(b) there are two types of contributions to these fixed points: along the $n_i-n_j=0$ - line and $n_i-n_j\neq 0$ with $\theta_i-\theta_j=-\pi/3\pm2\pi k$. The $n_i-n_j=0$ lines do not depend on the interaction, whereas the latter do. Thus this multidimensional fixed point is characterised just by two occupancies: $n$ and $n'$ ($n\neq n'$). The $n$ are the occupancies of the  outer triangles shown in white, gray and black colors in Fig. \ref{fig:EVk=1}(b), and the $n'$ are populations of the remaining sites. An equation that relates $n$, $n'$ and $u$ at stable fixed points can be readily found:
 \begin{equation}
     -u(n-n')+2\sqrt{\frac{n'}{n}}-\sqrt{\frac{n}{n'}}+1=0.
     \label{fp_u}
 \end{equation}
Using the normalisation condition $n+2n'=3$ we can express  the interaction at the fixed point $u^*$ in terms of $n'$
\begin{equation}
     \frac{1}{3(1-n')}\left(1+\frac{4n'-3}{\sqrt{(3-2n')n'}}\right)=u^*.
     \label{fp_u}
 \end{equation} 
The analysis of this equation shows that 
\begin{equation}
    \lim_{u^*\rightarrow \infty}n'=1_-, \quad \lim_{u^*\rightarrow \infty}n=1_+
\end{equation}
with the noninteracting values being $n'(u=0)=1/2$ and $n(u=0)=2$. This means that with increasing $u$ the fixed point is moving towards $n_i-n_j=0$, but never reaches it (we added some of the interaction-dependent fixed points to Fig. \ref{fig:periodic_dyn}(b) for $u=3$, $u=10$ and $u=13$). Due to these peculiarities, the trajectories around the fixed points in Fig. \ref{fig:periodic_dyn}(b) are never perfect ellipses, but rather egg-shaped.  

We note that in a bosonic Josephson junction of cold atoms all fixed points are stable for $u=0$. This is due to a supercritical pitchfork bifurcation  occurring  for certain values of $u$, that one of the fixed points changes its character to unstable \cite{Raghavan1999}. This change eventually leads to self-trapping. In contrast, our system is characterised by a multidimensional unstable fixed point, whose number of components does not change with the interaction. This fixed point can be effectively described with two components: one stable and one unstable, as can be clearly seen in the projected phase portrait in Fig. \ref{fig:periodic_dyn}. It is 
 the tendency of the stable component of the fixed point to align with the unstable one along $n_i-n_j=0$, a line with increasing interaction, which eventually causes the self-trapping.

\begin{figure}[h]
    \centering
    \includegraphics[width=0.8\linewidth, keepaspectratio]{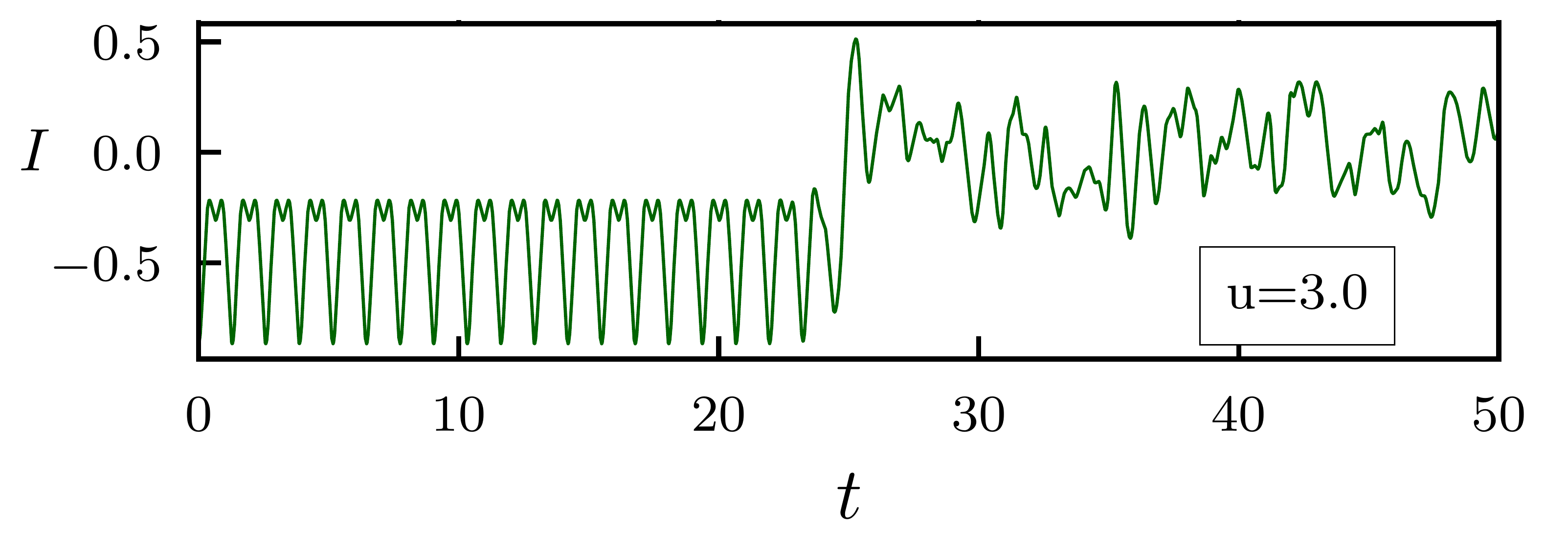}
     \caption{The total circular current $I$ (in units of $\rho K$) over time (in units of $1/K$) of a periodic mode for $u=3$, $k=2$. The dynamics changes from periodic to chaotic at $t_c\approx 25$. }
    \label{fig:per_chaos}
\end{figure}

Another difference with a bosonic Josephson junction, is that the oscillating dynamics of the periodic modes transforms after a while into chaotic. This is exemplified in Fig. \ref{fig:per_chaos}, where the total cyclic current is shown versus time for $u=3$. The characteristic time scale associated with chaotic dynamics $t_c$ is not easy to evaluate numerically due to obstructing oscillations. We therefore define a condition for the onset of chaos by first fitting the linear rise in amplitude from the upper part of Fig. \ref{fig:AmpltPlots}, resulting in: 
\begin{equation}
      I_{max}= 0.220(\pm 0.000) \cdot u - 0.878(\pm0.001).
\end{equation}
The value of the intercept should be the current value for the non-interacting case, $-\sqrt{3}/2 \approx -0.866$, the discrepancy could be attributed to the first few $I_{Max}$ values deviating from the linear curve more than any others. 

From these findings a condition for estimating $t_c$ can be formulated: if the current either exceeds the $I_{max}$ curve or goes below the minimum value by a certain margin (in this case $0.05$) the time is taken as the critical one for that $u$. This of course only applies to interaction values below $u=12.0$. The corresponding graph is presented in Fig. \ref{Fig:TC} (see the data for $\epsilon=-1$), where once again $t_c$ falls with rising interaction parameter; however, the $k=1$ case shows a pronounced kink. Since linear stability analysis is not applicable in this case, we can not make a conclusion about the origins of the kink. 

\subsection{Discussion of modes in higher generations}

In general all of the discussed energy levels will be present at any fractal stage $k\ge 2$. For this reason we can predict how the modes with the same energies will behave at $k=3$ and other stages. For example, standard modes with $\epsilon=2$ will become unstable for $u>u_c$ and will be characterised by the time-scale $t_c$ as well. However, $u_c$ as well as $t_c$ will depend on generation, with $u_c$ eventually tending to zero in the thermodynamic limit. As before, $t_c$ will be characterised by a kink; however, the number of those kinks in higher generations may be different from one, which  is, however, difficult to predict at this stage of work.  

We believe that periodic modes will show the same behaviour dependent on $u$ in all generations, as we saw no differences at least between the dynamics of $k=1$ and $k=2$ periodic modes for $u\neq 0$. 

We assume that edge modes which are characterised by isolated rings of circular current, should have chaotic dynamics similar to that of a simple three-site ring. However, we did not check that and this could be studied in future projects.

Modes with $\epsilon \approx -3.302$ at $k=2$ are generated by periodic modes with $\epsilon=-1$ at $k=1$, which in turn are generated by an $\epsilon =2$ state at $k=0$. Due to the recursion all these states will be present at all other stages, thus starting such a chain on all levels anew (one can also see this pattern from the schematics in Fig. \ref{fig:scheme}). Therefore the chaotic mode with quasi-localised bunches of points as in Fig. \ref{fig:chaotic_u}c) will be present for all $k\ge 2$. Modes with $\epsilon\approx 0.302$ are also gap states, however, they have differently structured state spaces, so that quasi-localised orbits do not appear in this modes for a finite interaction, contrary to the $\epsilon \approx -3.302$ gap states.

We suppose that for higher generations more and more chaotic modes will appear with so-called nice nonchaotic states being present only for energies $\epsilon \in \{-1,1,2\}$. 

\section{Conclusions and discussion}

\label{concl}

In this work we provided a detailed and systematic analysis of eigenmodes carrying loop currents in a system of weakly-interacting condensed bosons in Sierpinski gaskets. Essentially we explored the complex domain of discrete non-linear Schr\"odinger equations in fractal lattices. In a noninteracting system we identified energy levels which can accommodate such eigenstates  and found those states analytically where possible, or numerically. It turned out that the loop states can be very different in the distribution of their site populations as well as phase differences between nearest neighbors, often counter-intuitive. To make our main results more transparent, we divided the discovered modes into three main classes: standard, chaotic, and periodic. 

Standard modes correspond to the spectrum's upper boundary and are very similar in behaviour to the dynamics of the basic unit of the Sierpinski gasket, a simple three-site ring. They appear in all generations of the Sierpinski gasket and become unstable for certain finite values of $u$ for small gaskets with $u_c\rightarrow 0$ for larger systems. 

Chaotic loop modes can appear for a highly degenerate spectrum's upper boundary, as well for irrational eigenenergies (all energies expressed in units of coupling constant $K$). These modes appear first at the stage $k=2$ and we assume them to be typical of fractal lattices. One would not expect such modes in a regular lattice.

We also identifed modes, that we refer to as periodic. These modes are stationary states which are characterised by two contributions to the fixed point (due to symmetric phase space this suffices). One of them has zero occupation difference and remains where it is in the phase space even when the interaction is turned on. The other (for a different phase difference and nonzero population imbalance) changes the occupation difference with interaction strength. Thus the $u=0$ initial condition is no longer a fixed point and undergoes periodic dynamics. As we know from a simple pendulum, this kind of constellation of fixed points in phase space can lead to running phase modes, termed macroscopic self-trapping in the context of Bose Josephson junctions. This is indeed what happens for the periodic modes with increasing interaction. Interestingly, this behaviour due to self-similarity of the modes, is expected at all generations $k\ge 1$ of the Sierpinski gasket.

All in all, these are unexpected results, which should motivate further studies of these intriguing systems. For example, modes for $k=1$ and $\epsilon=1$ resemble edge modes in fractal photonic insulators \cite{Biesenthal2022} and deserve further exploration. 

Another line of research could be a generalisation to driven lattices, which is important because the driving force could account for the effect of gravitational field. At least for quasi-one-dimensional lattices of condensates the mean-field dynamics was in an excellent agreement with experiments probing full quantum many-body behaviour \cite{Kolovsky2009}. Since chaotic dynamics seems to be  a natural attribute of systems of coupled condensates, it would be instructive to measure its characteristics experimentally, for example, by Loschmidt echo methods, as suggested in \cite{Fine2017}. 

Finally, the role of quantum fluctuations and the study of possible thermalisation of bosons in fractal lattices could give researchers further clues about the role of chaos and eigenstate thermalisation hypothesis \cite{Deutsch1991,Srednicki1994,ETHreview2016}, as well as dynamical heat bath generation mechanisms \cite{Anna2016,Anna2018}.

\newpage

\section*{Appendix A: Spectrum of the noninteracting system}

Here we briefly outline the main results for the single-particle spectrum derived in Refs. \cite{Domany1983,Rammal1984}. The spectrum was calculated with the help of  a decimation procedure according to which the Hamiltonian of generation $k$ is mapped onto the Hamiltonian of the previous generation $k-1$. The procedure works consistently provided each site of a lattice is connected to the same number of nearest neighbors, in our case to the four nearest sites. In this case the basic structure of the effective Hamiltonian (the one defined on a subspace of "old" sites) does not change with on-site energies and effective coupling being renormalized in a simple way. 
To secure this requirement, 
periodic boundary conditions were suggested in \cite{Domany1983}, also shown in Fig. \ref{fig:fracpic}. This allowed  to calculate the spectrum of the infinite Sierpinski gasket. The spectrum is bounded, discrete and highly degenerate. 

As a result of the renormalization procedure, eigenenergies of generation $k$ fractal can be expressed recursively in terms of eigenenergies of the previous generation $k-1$
(in units of $K$)
\begin{equation}\label{eq:recursion}\epsilon_{k,\pm}=\frac{-3\pm\sqrt{(9-4\cdot\epsilon_{k-1})}}{2}.
\end{equation}
This recursion does not, however, account for all values of possible eigenenergies. There are three exceptional values $\epsilon^*=\pm 2; 1$, which should be treated separately (for these values the renormalization scheme fails, leading to either singularities or zeros in renormalized effective energies \cite{Domany1983}).  If we disregard the special values $\epsilon^*$ for the moment, we can analyse the stability of the recursive relation \eqref{eq:recursion},  and show that a sequence of spectral gaps is generated out of it in a regular manner. This sequence of gaps is in turn, a fractal described by a Cantor set of Lebesgue measure zero \cite{Domany1983,Rammal1984}.

\begin{figure}
    \centering
    \includegraphics[width=0.8\linewidth, keepaspectratio]{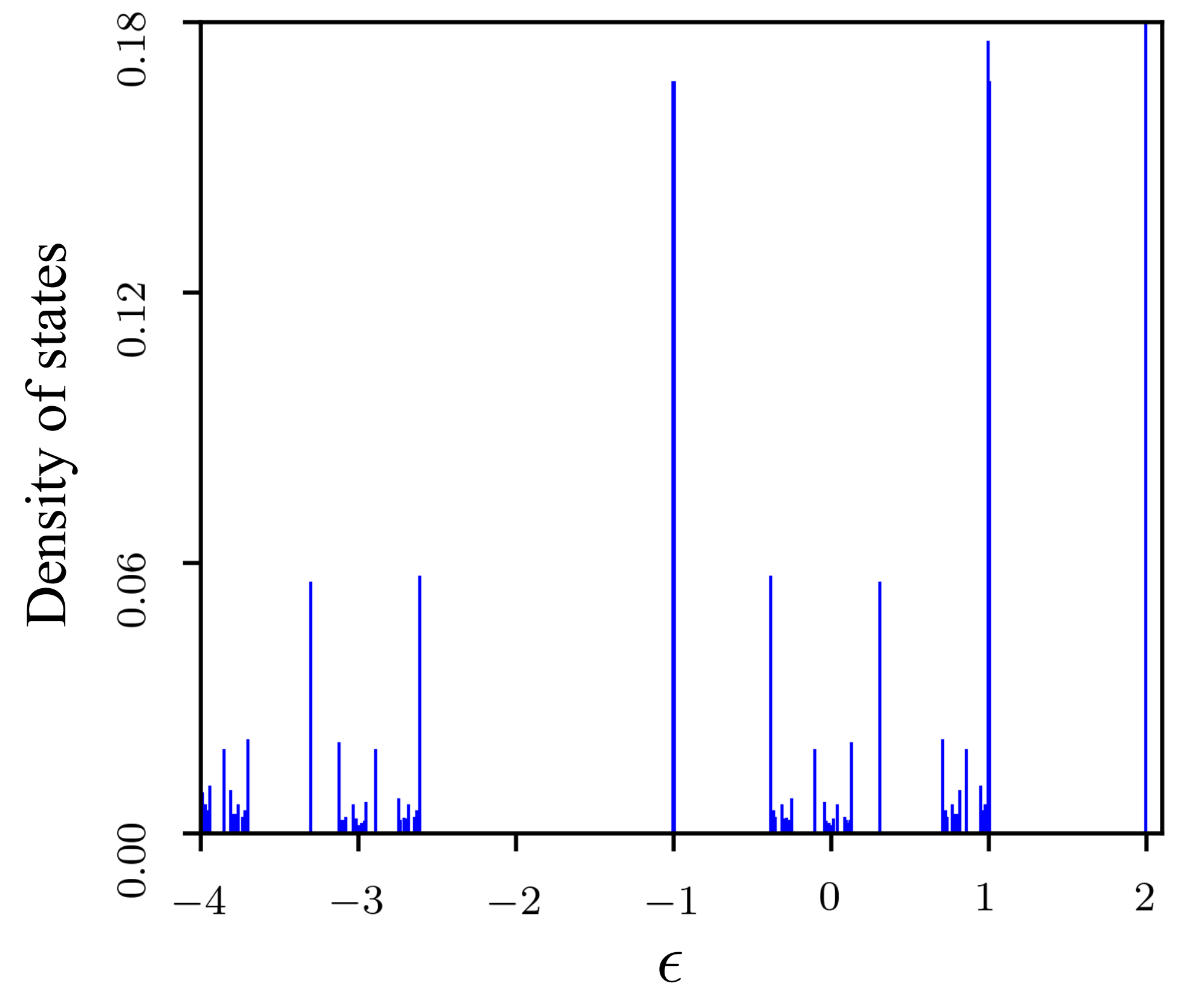}
    \caption{Density of states (DOS) for the first ten generations of the Sierpinski gasket. Such a DOS was first presented in \cite{Domany1983}. The peak at $\epsilon=2$ is very high due to the large degeneracy of the eigenvalue; we cut it at $0.18$. 
    }
    \label{fig:DOS}
\end{figure}

The whole spectrum as a result contains three parts: the special values, values inside the gap intervals, and values directly at the edges of the gaps. Values inside the gaps originate from $\epsilon=2$. Values at the gap edges descend from $\epsilon=1$. The entire spectrum is then a point set of Lebesgue measure zero. \cite{Domany1983}. 
The multiplicities of all values, special or not, are not difficult to calculate and as a result the density of states for the Sierpinski gasket of any size can be produced \cite{Domany1983}. For convenience, we reproduce here the density of states (DOS) for the first ten generations of the Sierpinski gasket (see Fig. \ref{fig:DOS}). We will refer to this DOS later in the text. Note, that the DOS exhibits self-similarity properties as well \cite{Domany1983}. 

 \section*{Appendix B: Details of hierarchical derivation of eigenvectors for $k=1$ and $k=2$ stages}

 The idea of hierarchical derivation was outlined in Ref. \cite{Domany1983}. Here we provide examples to show how the decimation procedure works  for stages $k=1$ and $k=2$.
 
 The eigenvalue problem for an eigenstate $|\Psi \rangle$ in generation $k$ is formulated in a two-dimensional space $\{|\Psi_1\rangle,|\Psi_2 \rangle \}$, where $|\Psi_1\rangle$ is the projection of $|\Psi \rangle$ onto the "old" subspace - the subspace of generation $k-1$, whereas $|\Psi_2\rangle$ is the projection of $|\Psi \rangle$ onto the "new" subspace, the one comprising all the "new" sites which were added
 in generation $k$. The Schr\"odinger equation can then be written in a convenient block-form
 \begin{equation}
    \begin{pmatrix}
         \mathcal{H}_{11} & \mathcal{H}_{12} \\
         \mathcal{H}_{21} & \mathcal{H}_{22} 
    \end{pmatrix}
    \begin{pmatrix}
         \ket{\psi_1}\\
         \ket{\psi_2}
    \end{pmatrix}
    =\epsilon
    \begin{pmatrix}
          \ket{\psi_1}\\
         \ket{\psi_2}
    \end{pmatrix}
    .
    \label{ham_block}
\end{equation}
Here $H_{11}$ is a block of zeros in the old subspace and  $H_{22}$ is a block-diagonal matrix of the size of the new subspace with $\hat S$ -matrices along the diagonal
\begin{equation}
\hat H_{22}=\begin{pmatrix} 
\hat S & 0 & 0  & ... & 0 \\
 0 & \hat S & 0 & ... & 0 \\
 ... & & & &  \\
 0 & 0 & 0 & ... & \hat S 
\end{pmatrix}, \quad 
\hat S=\begin{pmatrix*}[r]
         0& -1& -1 \\
         -1& 0 & -1 \\
         -1& -1& 0 \\
    \end{pmatrix*}.
\end{equation}
What changes from stage to stage is the form of  connecting block $H_{12}$ ($H_{21}=H_{12}^T$). 
For example, for stage $k=1$
\begin{equation}
    H_{12}=(H_{21})^T=(\hat S\, \hat S).
\end{equation}
In the next stages $H_{12}$ will contain similar matrices, whose sum per row will be equal to $2\hat S$, and so on (see  Eeq. \eqref{H21_k_2}).

 We can now express the new eigenvectors $\ket{\psi_2}$ in terms of the old states $\ket{\psi_1}$
\begin{equation}
    \ket{\psi_2}= (\epsilon\, \hat{\mathbb{I}}-\mathcal{H}_{22})^{-1}\mathcal{H}_{21}\ket{\psi_1}, 
    \label{eq:psi2}
\end{equation}
provided $\epsilon\, \hat{\mathbb{I}}-\mathcal{H}_{22}$ is invertible. 
Since $H_{22}$ is block-diagonal, it is easy to invert the matrix $(\epsilon\, \hat{\mathbb{I}}-\mathcal{H}_{22})$ just by inverting each block 
\begin{eqnarray}
    (\epsilon\, \hat{\mathbb{I}}-\hat S)^{-1}&=&\frac{1}{\epsilon^2+\epsilon-2}\begin{pmatrix} 1+\epsilon & -1 & -1 \\
    -1 & 1+ \epsilon & -1 \\ -1 & -1 & 1+ \epsilon \end{pmatrix} = \nonumber \\
   & =&\frac{1}{\epsilon^2+\epsilon-2}((1+\epsilon) \hat{\mathbb{I}}+\hat S)\equiv \hat B. 
   \label{eq:inverse}
\end{eqnarray}
We see that inversion does not work for  $\epsilon=1$ and $\epsilon=-2$. These cases are to be treated separately. However, $\epsilon=-2$ is not a loop current state and will not be considered here.

Consider, for example, $k=1$. In this generation we can calculate an eigenvector, generated by $\ket{\psi_1}=V$ defined in  Eq. \eqref{eq:V}. Here $V$ corresponds to the eigenvalue $\epsilon=2$ in generation $k=0$.  In generation $k=1$ this eigenvalue will generate $\epsilon=-1$ (see the flowchart in Fig. \ref{fig:EE}); therefore, we insert $\epsilon=-1$ and $V$ instead of $|\psi_1\rangle$ in Eqs.  \eqref{eq:psi2} and \eqref{eq:inverse}.  After some algebra we get 
\begin{equation}
    (\epsilon\, \hat{\mathbb{I}}-\mathcal{H}_{22})^{-1}\mathcal{H}_{21}=-\begin{pmatrix} 1 & 0.5 & 0.5 \\ 0.5 & 1 & 0.5 \\ 0.5 & 0.5 & 1 \\ 1 & 0.5 & 0.5 \\ 0.5 & 1 & 0.5 \\ 0.5 & 0.5 & 1 \end{pmatrix}.
    \label{eq:psi2algebra}
\end{equation}
Now from Eq. \eqref{eq:psi2} we get
\begin{equation}
   \ket{\psi_2}=\begin{pmatrix}  -1/2 \\ (-i\sqrt{3}+1)/4 \\ (i\sqrt{3}+1)/4 \\ -1/2 \\ (-i\sqrt{3}+1)/4 \\ (i\sqrt{3}+1)/4 \end{pmatrix}  = \begin{pmatrix} -0.5 \\ 0.5 e^{-i \pi/3} \\ 0.5 e^{i \pi/3} \\ -0.5 \\ 0.5 e^{-i \pi/3} \\ 0.5 e^{i \pi/3} \end{pmatrix},
\end{equation}
so that 
\begin{equation}
    |\Psi \rangle =  \begin{pmatrix}
         \ket{\psi_1}\\
         \ket{\psi_2}
    \end{pmatrix}= \begin{pmatrix} 1 \\ e^{i 2\pi/3} \\ e^{i 4 \pi/3} \\-0.5 \\ 0.5 e^{-i \pi/3} \\ 0.5 e^{i \pi/3} \\ -0.5 \\ 0.5 e^{-i \pi/3} \\ 0.5 e^{i \pi/3} \end{pmatrix}
\end{equation}
which up to a factor is the eigenvector $|\Psi_{k=1}(-1) \rangle$ in Eq. \eqref{eq:EVk=1}. Thus in generation $k=1$ only one loop current state can be generated from the previous stage. The other two are for special values $\epsilon=1$ and $\epsilon=2$ and  should be found directly from the Schr\"odinger equation. 

\begin{figure}[h]
    \centering   \includegraphics[width=0.6\linewidth, keepaspectratio]{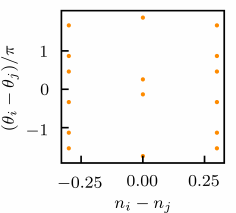}
    \caption{Resulting state space from a linear combination of all the discussed chaotic modes ($\epsilon=2$, and $k=2$). For the one seen in Fig. \ref{fig:EVk=2_eps2} b) the complex conjugation was applied beforehand. The mode is now single-colored, because the three subtriangles are equivalent in contrast to Fig.\ref{fig:EVk=2_eps2} a) and b).}
    \label{fig:SymmMode}
\end{figure}

In the next stage $k=2$ the partitioned Hamiltonian has the following blocks: $H_{11}$ is a $9\times 9$ matrix of zeros, $H_{22}$ is a $6\times 6 $ block-diagonal matrix with blocks $\hat S$ on its main diagonal, so that 
$   (\epsilon\, \hat{\mathbb{I}}-\mathcal{H}_{22})^{-1}$ is a block-diagonal matrix with each block equal to $\hat B$ from Eq. \eqref{eq:inverse}. For $H_{21}$ we get
\begin{equation}
    H_{21}=-\begin{pmatrix} \hat B_1 & \hat B_2 & \hat 0 \\ \hat B_3 & \hat B_4 & \hat 0 \\ \hat B_5 & \hat B_6 & \hat 0 \\ \hat B_1 & \hat 0&  \hat B_2 \\ \hat B_3 &  \hat 0 & \hat B_4  \\ \hat B_5 &  \hat 0 & \hat B_6 \end{pmatrix},
    \label{H21_k_2}
    \end{equation}
    where
    \begin{eqnarray}
        \hat B_1= \begin{pmatrix}
        0 & 0 & 0 \\ 1 & 0 & 0 \\ 1 & 0 & 0 \end{pmatrix},\, \hat B_2= \begin{pmatrix}
        0 & 1 & 1 \\ 0 & 0 & 1 \\ 0 & 1 & 0 \end{pmatrix}, \nonumber \\
         \hat B_3= \begin{pmatrix}
        0 & 1 & 0 \\ 0 & 0 & 0 \\ 0 & 1 & 0 \end{pmatrix},\, \hat B_4= \begin{pmatrix}
        0 & 0 & 1 \\ 1 & 0 & 1 \\ 1 & 0 & 0 \end{pmatrix}, \nonumber \\
         \hat B_5= \begin{pmatrix}
        0 & 0 & 1 \\ 0 & 0 & 1 \\ 0 & 0 & 0 \end{pmatrix},\, \hat B_6= \begin{pmatrix}
        0 & 1 & 0 \\ 1 & 0 & 0 \\ 1 & 1 & 0 \end{pmatrix},  
            \end{eqnarray}
            with properties $-\hat B_1-\hat B_3-\hat B_5=\hat S$ and $-\hat B_2-\hat B_4-\hat B_6=2\hat S$. We get then 
            \begin{equation}
  (\epsilon\, \hat{\mathbb{I}}-\mathcal{H}_{22})^{-1}\mathcal{H}_{21}=-\begin{pmatrix} \hat B\cdot \hat B_1 &  \hat B\cdot \hat B_2 & 0 \\  \hat B\cdot \hat B_3 &  \hat B\cdot \hat B_4 & 0 \\ \hat B\cdot \hat B_5 &  \hat B\cdot \hat B_6 & 0  \\ \hat B\cdot \hat B_1 & 0 &  \hat B\cdot \hat B_2 \\  \hat B\cdot \hat B_3 & 0 &  \hat B\cdot \hat B_4 \\ \hat B\cdot \hat B_5 & 0 &  \hat B\cdot \hat B_6 \end{pmatrix}.
  \label{eq:matrix}
\end{equation}
Now we can plug  $|\psi_1\rangle =(V,V,V)^T$ and $\epsilon=2$ in Eqs.  \eqref{eq:psi2} and \eqref{eq:inverse} and derive 
$|\Psi_2\rangle= (V,V,V,V,V,V)^T$. This is to say that the standard eigenvector of the type depicted in Fig. \ref{fig:EVk=1}(a) will be present in generation $k=2$ and in fact, in all generations for $\epsilon=2$. However, $\epsilon=2$ is a highly degenerate state and the remaining three eigenvectors can be calculated only numerically. Here we present only the $|\Psi_2\rangle$ part of them in order  to save space (the vectors are then appended by $|\Psi_1\rangle=(V V V)^T$):
\begin{widetext}

    \begin{equation}
\ket{\psi_2(\epsilon=2)}_{2}=
\begin{pmatrix}
    1.440\\
    0.5 + 0.342\cdot i\\
    -0.5 - 0.342\cdot i\\
    -0.163 + 0.283\cdot i\\
    1.103 + 0.059\cdot i\\
    -0.603 - 0.925\cdot i\\
    0.133 - 0.230\cdot i\\
    -0.307 - 0.754\cdot i\\
    0.807 - 0.112\cdot i\\
    0.030 + 0.814 \cdot i\\
    -0.796 - 0.171 \cdot i\\
    -0.204 + 0.171 \cdot i \\
    -1.016 - 0.433 \cdot i \\
     0.25 + 1.076 \cdot i \\
     0.25 - 0.210 \cdot i\\
     -0.424 - 0.433 \cdot i\\
      0.25 - 0.552 \cdot i\\
      0.25 + 1.418 \cdot i
\end{pmatrix},  \ket{\psi_2(\epsilon=2)}_{3}=
\begin{pmatrix}
   -0.266\\
   -0.5 - 0.643\cdot i\\
   -0.5 + 0.643\cdot i\\
   -0.720 - 1.247\cdot i\\
   -0.046 + 0.604\cdot i\\
   0.546 + 0.262\cdot i\\
   -0.163 - 0.283\cdot i\\
   1.103 - 0.059\cdot i\\
   -0.603 + 0.925 \cdot i\\
   0.883 + 0.663 \cdot i\\
   -1.057 + 0.321 \cdot i\\
   0.057 - 0.321 \cdot i\\
   -0.424 + 0.433 \cdot i\\
   0.25 + 0.552 \cdot i\\
   0.25 - 1.418 \cdot i\\
   0.690 + 0.433 \cdot i\\
   0.25 - 0.775 \cdot i\\
   0.25 - 0.091 \cdot i\\
\end{pmatrix},  
\ket{\psi_2(\epsilon=2)}_{4}=
\begin{pmatrix}
    0.326\\
    -0.5 - 0.985 \cdot i\\
    -0.5 + 0.985 \cdot i\\
    0.133 - 0.230 \cdot i\\
    -0.307 - 0.754 \cdot i\\
    0.807 - 0.112 \cdot i\\
    -0.720 + 1.247 \cdot i\\
    -0.046 - 0.604 \cdot i\\
    0.546 - 0.262 \cdot i\\
    0.587 - 0.150 \cdot i\\
    0.353 + 0.492 \cdot i\\
    -1.353 - 0.492 \cdot i\\
    0.690 - 0.433 \cdot i\\
    0.25 + 0.775 \cdot i\\
    0.25 + 0.091 \cdot i\\
   -1.016 - 0.433 \cdot i\\
    0.25 + 1.076 \cdot i \\
    0.25 - 0.210 \cdot i\\
\end{pmatrix}. \nonumber
\end{equation}    

\end{widetext}
Their state space representation looks chaotic. A linear combination of them ($|\psi_2\rangle_2+|\psi_2\rangle_3^*+|\psi_2\rangle_4$) results in a symmetric state space, shown in Fig. \ref{fig:SymmMode}. 

We now discuss periodic modes of stage $k=2$. These are loop current eigenstates for $\epsilon=-1$, which are generated from the standard modes of the previous generation shown in  Fig. \ref{fig:EVk=1} (a).  In order to calculate the new modes we put $\epsilon=-1$ and $|\psi_1\rangle =(V, V, V)^T$  in Eq. \eqref{eq:psi2}.  When we multiply the matrix of Eq. \eqref{eq:matrix} with  $|\psi_1\rangle =(V, V, V)^T$ we get a very simple result
\begin{equation}
    |\psi_2\rangle = - \frac{1}{2}(V,V,V,V,V,V)^T,
\end{equation}
because 
\begin{eqnarray}
    \hat B\cdot \hat B_1 + \hat B\cdot \hat B_2&=&\hat B\cdot \hat B_3 + \hat B\cdot \hat B_4=\hat B\cdot \hat B_5 + \hat B\cdot \hat B_6=\nonumber \\ &-&\frac{1}{2}\hat S (\hat B_1+\hat B_2)=\frac{1}{2}\hat S^2. 
\end{eqnarray}
These eigenmodes hence preserve the structure of the periodic modes in Fig. \ref{fig:EVk=1}(b) with old subspace having larger occupation numbers and phase shifted by $\pi$ with respect to triangles of the new subspace. These structure will be preserved in all generations.  

We proceed to loop states for the eigenvalues $\epsilon\approx -0.381$ and $\epsilon \approx -2.618$. According to the flowchart in Fig. \ref{fig:EE} they are derived from partially filled states of generation $k=1$, which means from $|\psi_1\rangle =(0, -V, V)^T$. This leads to 
\begin{equation}
    |\Psi_2\rangle =\begin{pmatrix} \hat B \cdot\hat B_2 \cdot V\\  \hat B \cdot\hat B_4 \cdot V \\  \hat B \cdot\hat B_6 \cdot V \\ -\hat B \cdot\hat B_2 \cdot V\\  -\hat B \cdot\hat B_4 \cdot V \\  -\hat B \cdot\hat B_6 \cdot V\end{pmatrix}. 
\end{equation}
Since, for example,  
\begin{equation}
    \hat B \cdot\hat B_2 \cdot V=\begin{pmatrix}  -\epsilon \\ -2e^{i2\pi/3}+\epsilon\, e^{-i2\pi/3} \\ -2e^{-i2\pi/3}+\epsilon\, e^{i2\pi/3} \end{pmatrix}
\end{equation}
 and $\epsilon$ is an irrational number, the entries of this vector can not be simplified to a "nice" Euler form, with argument being just a rational fraction of $\pi$. What we see is that the sites which were unfilled  in Fig. \ref{fig:EVk=1}(c)) remain unfilled, thus effectively splitting the fractal into two independent subspaces.

Next loop current states, corresponding to $\epsilon\approx 0.302$ and $\epsilon\approx -3.302$ are derived from $|\psi_1\rangle=(-2V,V,V)^T$. so that  
\begin{equation}
    |\Psi_2\rangle =\begin{pmatrix} (2\hat B \cdot\hat B_1- \hat B \cdot\hat B_2) V\\  (2\hat B \cdot\hat B_3-\hat B \cdot\hat B_4) V \\ ( 2\hat B \cdot\hat B_5-\hat B \cdot\hat B_6)V \\  (2\hat B \cdot\hat B_1-\hat B \cdot\hat B_2) V\\  (2\hat B \cdot\hat B_3-\hat B \cdot\hat B_4) V \\ (2\hat B \cdot\hat B_5-\hat B \cdot\hat B_6)V \end{pmatrix},
    \label{eq:psi2case}
\end{equation}
which is again difficult to simplify due to irrational character of the eigenenergies (blocks $\hat B$ are all $\epsilon$-dependent). We therefore resume to numerics and get (we only present the upper halves of the vectors, since the lower parts are the same as is clear from Eq. \eqref{eq:psi2case}):
%\begin{widetext}
\begin{eqnarray}
\ket{\psi_2(\epsilon\approx -3.302)}= 
\begin{pmatrix}
     0.651\\
     0.826 + 0.101 \cdot i \\
     0.826 - 0.101 \cdot i \\
     -0.326 + 0.765 \cdot  i \\
     -0.326 + 0.564 \cdot  i \\
     -0.5 + 0.665 \cdot i \\
     -0.326 - 0.765 \cdot  i \\
     -0.5 - 0.665 \cdot i \\
     -0.326 - 0.564 \cdot  i 
      \end{pmatrix}, \nonumber \\  \ket{\psi_2(\epsilon\approx 0.302)}= 
\begin{pmatrix}
   -1.151\\
   -0.076 + 0.621 \cdot i\\
   -0.076 - 0.621 \cdot i\\
    0.576 + 0.245 \cdot i\\
    0.576 - 0.997 \cdot i \\
    -0.5 - 0.376 \cdot i \\
    0.576 - 0.245 \cdot i\\
   -0.5 + 0.376 \cdot i \\
   0.576 + 0.997 \cdot i 
  \end{pmatrix}.
  \label{eq:quasi_loc}
\end{eqnarray}   
%\end{widetext}

The last loop current states for generation $k=2$ to be discussed are the "edge" states, corresponding to $\epsilon=1$, the band edge energy. These states are not possible to derive from Eqs. \eqref{eq:psi2} and \eqref{eq:inverse}. However, one can see that the global structure of the states $(0,-V,V)$ will be preserved in all generations. For example, in generation $k=2$ this vector will be
\begin{equation}
    (\hat 0,\hat 0,\hat 0, \hat v_1, \hat v_2, \hat v_3, -\hat v_1,-\hat v_2,-\hat v_3)^T
    \label{condition}
\end{equation}
This means, that the old subspace will be empty in each generation, whereas the new subspace will be filled (albeit partially as we see below). 

The three vectors $\hat v_i$, which we need to find, are three-dimensional vectors, which satisfy the equation
\begin{equation}
    -\hat B_2^T\hat v_1-\hat B_4^T\hat v_2-\hat B_6^T\hat v_3=\hat 0. 
\end{equation}
This equation follows from Eq. \eqref{ham_block}. Moreover, another property of $\hat v_i$-vectors, which also follows from Eq.  \eqref{ham_block} is that they are all eigenvectors of the $\hat S$-matrix, corresponding to eigenvalue $\epsilon=1$. These eigenvectors of $\hat S$ can be expressed in many different ways, e.g. the three real ones can be found as  
\begin{equation}
    \hat v_1^r=\begin{pmatrix}
        0 \\ -1 \\ 1
    \end{pmatrix}, \quad \hat v_2^r=\begin{pmatrix}
        1 \\ 0 \\ -1
    \end{pmatrix}, \quad \hat v_3^r=\begin{pmatrix}
        -1 \\ 1 \\ 0
    \end{pmatrix}. 
    \label{real_eigenvectors}
\end{equation}
Since $\hat B_2, \hat B_4$ and $\hat B_6$ are similar matrices, they all share an eigenvalue $\epsilon=1$. Eigenvectors, corresponding to this eigenvalue are of course different, but they happen to coincide with the vectors in \eqref{real_eigenvectors}. In this case the condition \eqref{condition} is trivially satisfied. 
This is unfortunately not the case for complex eigenvectors, because matrices $\hat B_{2,4,6}$ do not share complex eigenvectors with $\hat S$. 

After some effort we get
\begin{equation}
    \hat v_1=\begin{pmatrix}
        e^{-i2\pi/3} \\ e^{i2\pi/3} \\ 1
    \end{pmatrix}, \quad \hat v_2=\hat v_2^r, \quad \hat v_3=-\begin{pmatrix}
        1  \\ e^{i2\pi/3} \\ e^{-i 2\pi/3}
    \end{pmatrix},
\end{equation}
which means only two (per subspace) loop currents can be maintained.

\section*{Appendix C: Linear fits of logarithmic plot in Fig. \ref{fig:TC_LOG} }\label{App:c}

Tables \ref{slopes1} and \ref{slopes2} contain the slopes (the values of $a$ from Eq. \eqref{eq:linear_fit}) from the linear fit of $\ln(t_c)$ versus $\ln(u-u_c)$. 
It is difficult to rigorously determine the position of the kink in  the graphs. For $k=1$ the kink is around $\ln (u-u_c)\approx 0$. For $k=2$ it is shifted to smaller values of $\ln(u-u_c)\approx -1$. In Tables \ref{slopes1} and \ref{slopes2} we present the tabulated values of $a$ for two linear segments: segment I before the kink, and segment II after the kink. 

\begin{table}[h]
\caption{Slopes $a$ from the linear fittings of the doubly logarithmic plots in Fig. \ref{fig:TC_LOG} for $k=1$. }
\begin{tabularx}{3.48in}{||>{\centering\arraybackslash}X|>
{\centering\arraybackslash}X|>
{\centering\arraybackslash}X||}
\hline
$a$ in segment I & $a$ in segment II & $-\log{\epsilon}$  \\ 
\hline
 $-0.303(\pm 0.004)$ & $-0.745(\pm 0.001)$  &  $0$ \\
 $-0.367(\pm 0.004)$ & $-0.697(\pm 0.002)$  &  $1$ \\
 $-0.396(\pm 0.004)$ & $-0.778(\pm 0.004)$  &  $2$ \\
 $-0.412(\pm 0.004)$ & $-0.802(\pm 0.002)$  &  $3$ \\
 $-0.422(\pm 0.004)$ & $-0.750(\pm 0.001)$  &  $4$ \\ 
 $-0.429(\pm 0.004)$ & $-0.705(\pm 0.002)$  &  $5$ \\
 $-0.409(\pm 0.005)$ & $-0.714(\pm 0.004)$  &  $6$ \\
 $-0.413(\pm 0.005)$ & $-0.746(\pm 0.004)$  &  $7$ \\
 $-0.395(\pm 0.006)$ & $-0.760(\pm 0.003)$  &  $8$ \\
 $-0.397(\pm 0.006)$ & $-0.759(\pm 0.002)$  &  $9$ \\
 $-0.382(\pm 0.006)$ & $-0.751(\pm 0.002)$  &  $10$ \\
 $-0.383(\pm 0.006)$ & $-0.747(\pm 0.002)$  &  $11$ \\
 $-0.369(\pm 0.007)$ & $-0.741(\pm 0.002)$  &  $12$ \\  
 \hline
\end{tabularx}
\label{slopes1}
\end{table}

\begin{table}[h]
\caption{Slopes $a$ from the linear regression of the doubly logarithmic plots in Fig. \ref{fig:TC_LOG} for $k=2$. }
\begin{tabularx}{3.48in}{||>{\centering\arraybackslash}X|>
{\centering\arraybackslash}X|>
{\centering\arraybackslash}X||}
\hline
$a$ in segment I & $a$ in segment II & $-\log{\epsilon}$  \\ 
\hline
$-0.372(\pm 0.021)$ &  $-0.950(\pm 0.005)$ & $0$ \\
$-0.360(\pm 0.010)$ &  $-0.942(\pm 0.003)$ & $1$ \\
$-0.358(\pm 0.005)$ &  $-0.924(\pm 0.003)$ & $2$ \\
$-0.330(\pm 0.006)$ &  $-0.907(\pm 0.003)$ & $3$ \\
$-0.302(\pm 0.007)$ &  $-0.892(\pm 0.002)$ & $4$ \\
$-0.278(\pm 0.007)$ &  $-0.882(\pm 0.002)$ & $5$ \\
$-0.259(\pm 0.007)$ &  $-0.872(\pm 0.002)$ & $6$ \\
$-0.243(\pm 0.008)$ &  $-0.860(\pm 0.002)$ & $7$ \\
$-0.215(\pm 0.008)$ &  $-0.840(\pm 0.001)$ & $8$ \\
$-0.196(\pm 0.009)$ &  $-0.821(\pm 0.001)$ & $9$ \\
$-0.185(\pm 0.011)$ &  $-0.805(\pm 0.001)$ & $10$ \\
$-0.179(\pm 0.013)$ &  $-0.791(\pm 0.001)$ & $11$ \\
$-0.178(\pm 0.016)$ &  $-0.777(\pm 0.001)$ & $12$ \\ 
\hline
\end{tabularx}
\label{slopes2}
\end{table}

\newpage

\bibliographystyle{acm}
\bibliography{bibliography.bib}

\end{document}